\documentclass[twocolumn]{aastex62}

\newcommand{\mygal}{A1689-217}
\newcommand{\redshift}{2.5918}

\begin{document}
\title{THE DETECTION OF [\ion{O}{3}]$\lambda$4363 IN A LENSED, DWARF GALAXY AT $z$ = 2.59: TESTING METALLICITY INDICATORS AND SCALING RELATIONS AT HIGH REDSHIFT AND LOW MASS\footnote{The data presented herein were obtained at the W. M. Keck Observatory, which is operated as a scientific partnership among the California Institute of Technology, the University of California and the National Aeronautics and Space Administration. The Observatory was made possible by the generous financial support of the W. M. Keck Foundation. \\
Based on observations made with the NASA/ESA Hubble Space Telescope, obtained from the Data Archive at the Space Telescope Science Institute, which is operated by the Association of Universities for Research in Astronomy, Inc., under NASA contract NAS5-26555. These observations are associated with programs $\#$9289, $\#$11710, $\#$11802, $\#$12201, $\#$12931.}}

\correspondingauthor{Timothy Gburek}
\email{timothy.gburek@email.ucr.edu}

\author{Timothy Gburek}
\affiliation{Department of Physics \& Astronomy, University of California, Riverside, CA 92521, USA}

\author{Brian Siana}
\affiliation{Department of Physics \& Astronomy, University of California, Riverside, CA 92521, USA}

\author{Anahita Alavi}
\affiliation{Infrared Processing and Analysis Center, Caltech, Pasadena, CA 91125, USA}
\affiliation{Department of Physics \& Astronomy, University of California, Riverside, CA 92521, USA}

\author{Najmeh Emami}
\affiliation{Department of Physics \& Astronomy, University of California, Riverside, CA 92521, USA}

\author{Johan Richard}
\affiliation{Univ Lyon, Univ Lyon1, Ens de Lyon, CNRS, Centre de Recherche Astrophysique de Lyon UMR5574, F-69230, Saint-Genis-Laval, France}

\author{William R. Freeman}
\affiliation{Department of Physics \& Astronomy, University of California, Riverside, CA 92521, USA}

\author{Daniel P. Stark}
\affiliation{Department of Astronomy, Steward Observatory, University of Arizona, 933 North Cherry Avenue, Rm N204, Tucson, AZ 85721, USA}

\author{Christopher Snapp-Kolas}
\affiliation{Department of Physics \& Astronomy, University of California, Riverside, CA 92521, USA}

\author{Breanna Lucero}
\affiliation{Department of Physics \& Astronomy, University of California, Riverside, CA 92521, USA}

\begin{abstract}
    \noindent We present Keck/MOSFIRE and Keck/LRIS spectroscopy of \mygal, a lensed (magnification $\sim7.9$), star-forming (SFR $\sim$ 16 $\rm M_\sun\ yr^{-1}$), dwarf (log($M_\ast/M_\sun$) = $8.07-8.59$) Ly$\alpha$-emitter ($\rm EW_0\sim138$ $\rm \AA$) at $z$ = \redshift. Dwarf galaxies similar to \mygal\ are common at high redshift and likely responsible for reionization, yet few have been studied with detailed spectroscopy. We report a 4.2$\sigma$ detection of the electron-temperature-sensitive [\ion{O}{3}]$\lambda$4363 emission line and use this line to directly measure an oxygen abundance of 12+$\log$(O/H) = 8.06 $\pm$ 0.12 ($\sim1/4\ \rm Z_\sun$). \mygal\ is the lowest mass galaxy at $z>$ 2 with an [\ion{O}{3}]$\lambda$4363 detection. Using the rest-optical emission lines, we measure \mygal's other nebular conditions including electron temperature ($T_e$([\ion{O}{3}]) $\sim$ 14,000 K), electron density ($n_e\sim$ 220 $\rm cm^{-3}$) and reddening ($E(\bv)\sim$ 0.39). We study relations between strong-line ratios and direct metallicities with \mygal\ and other galaxies with [\ion{O}{3}]$\lambda$4363 detections at $z\sim0-3.1$, showing that the locally-calibrated, oxygen-based, strong-line relations are consistent from $z\sim0-3.1$. We also show additional evidence that the $\rm O_{32}$ vs. $\rm R_{23}$ excitation diagram can be utilized as a redshift-invariant, direct-metallicity-based, oxygen abundance diagnostic out to $z\sim$ 3.1. From this excitation diagram and the strong-line ratio $-$ metallicity plots, we observe that the ionization parameter at fixed O/H is consistent with no redshift evolution. Although \mygal\ is metal-rich for its $M_\ast$ and SFR, we find it to be consistent within the large scatter of the low-mass end of the Fundamental Metallicity Relation.
\end{abstract}


\section{Introduction} \label{sec:intro}        
Gas-phase metallicity, measured as nebular oxygen abundance, is a fundamental property of galaxies and is critical to understanding how they evolve across cosmic time. Metallicity traces the complex interplay between heavy element production via star formation/stellar nucleosynthesis and galactic gas flows, whereby infalling gas dilutes the interstellar medium (ISM) with metal-poor gas, and outflowing gas removes metals from the galaxy. These gas flows also relate to star formation and feedback, in which cold gas falls into the galaxy, triggering star formation that is later quenched by enriched outflows from supernovae that heat the ISM and remove the gas needed for star formation. As a tracer of the history of inflows and outflows, metallicity measurements at different redshifts constrain the timing and efficiency of processes responsible for galaxy growth. 

This connection between metallicity and the build-up of stellar mass is encapsulated in the stellar mass ($M_\ast)$ $-$ gas-phase metallicity ($Z$) relation (MZR) of star-forming galaxies, seen both locally \citep[e.g.,][]{Tremonti2004,Kewley&Ellison2008,Andrews&Martini2013} and at high redshift \citep[e.g.,][]{Erb2006,Maiolino2008,Zahid2013,Henry2013,Steidel2014,Sanders2015,Sanders2019} where metallicities are lower at fixed stellar mass. The relation shows that low-mass galaxies are more metal-poor than their high-mass counterparts, possibly due to the increased effectiveness of galactic outflows (feedback) in shallower potential wells. Constraining the MZR and its redshift evolution is vital to constraining the processes ultimately responsible for galaxy formation and evolution.

The mass-metallicity relation has also been shown to derive from a more general relation between stellar mass, star formation rate (SFR), and oxygen abundance. This $M_\ast-\rm SFR-\textit{Z}$ connection, the Fundamental Metallicity Relation (FMR), was first shown to exist by \citet{Mannucci2010} with $\sim$140,000 Sloan Digital Sky Survey \citep[SDSS;][]{Abazajian2009} galaxies, and independently by \citet{Lara-Lopez2010} with $\sim$33,000 SDSS galaxies. The FMR constitutes a 3D surface with these three properties, for which metallicity is tightly dependent on stellar mass and SFR with a residual scatter of $\sim$0.05 dex \citep{Mannucci2010}, a reduction in the scatter observed in the MZR. The FMR is also observed to be redshift-invariant out to $z = 2.5$ \citep[][see also sources within the review of \citealt{Maiolino&Mannucci2019}]{Mannucci2010}, suggesting that the observed evolution of the MZR over this redshift range is the result of observing different parts of the locally-defined FMR at different redshifts. Above $z$ = 2.5, galaxies have lower metallicities than predicted by the locally-defined FMR \citep{Mannucci2010,Troncoso2014,Onodera2016}. These studies analyze galaxies at $z\gtrsim3$, where the strong optical emission lines used for metallicity determination are again observable in the H-band and K-band.

To accurately constrain the evolution of the MZR and FMR across redshift, metallicities must be estimated via a method that is consistent at all redshifts. Ideally, this is accomplished through first measuring other intrinsic nebular properties that dictate the strength of the collisionally-excited emission lines necessary for oxygen abundance determination. This ``direct" method estimates the electron temperature ($T_e$) and density ($n_e$) of nebular gas, in conjunction with flux ratios of strong oxygen lines to Balmer lines, to determine the total oxygen abundance \citep[e.g.,][]{Izotov2006}. Electron temperature is calculated via a temperature-sensitive ratio of strong emission lines, commonly [\ion{O}{3}]$\lambda$5007, to auroral emission lines, such as [\ion{O}{3}]$\lambda$4363 or \ion{O}{3}]$\lambda\lambda$1661,1666, from the same ionic species. The [\ion{O}{3}]$\lambda$4363 line and flux ratio of [\ion{O}{3}]$\lambda\lambda$4959,5007/[\ion{O}{3}]$\lambda$4363 is preferred as all lines lie in the rest-optical part of the electromagnetic spectrum. However, the [\ion{O}{3}]$\lambda$4363 line is faint, $\sim40-100$ times weaker than [\ion{O}{3}]$\lambda$5007 in low, sub-solar metallicity galaxies, and still weaker in higher-metallicity sources where metal cooling is more efficient. This makes observing the line difficult locally, and especially difficult at high redshift. Only 11 galaxies at $z > 1$ have been detected (most via gravitational lensing) with significant [\ion{O}{3}]$\lambda$4363 \citep{Yuan&Kewley2009,Brammer2012,Christensen2012,Stark2013,James2014,Maseda2014,Sanders2016o3,Sanders2019}, and of those only 3 are at $z>2$ \citep{Sanders2016o3,Sanders2019}.

In an effort to circumvent this problem and extend our ability to measure oxygen abundance to both high-metallicity and high-redshift galaxies, ``strong-line" methods were developed to estimate abundances via flux ratios of strong, nebular emission lines \citep[e.g.,][]{Jensen1976,Alloin1979,Pagel1979,Storchi-Bergmann1994}. These indirect methods utilize calibrations of the correlations between these strong-line ratios and metallicities derived empirically with direct metallicity measurements of nearby \ion{H}{2} regions and galaxies \citep[e.g.,][]{Pettini&Pagel2004,Pilyugin&Thuan2005}, theoretically with photoionization models \citep[e.g.,][]{McGaugh1991,Kewley&Dopita2002,Dopita2013}, or with a combination of both \citep[e.g.,][]{Denicolo2002}. However, as almost all of these calibrations have been done locally due to the inherent observational difficulties of the $T_e$-based, direct method (see \citealt{Jones2015} for the first calibrations done at an appreciable redshift, $z\sim$ 0.8), the question has naturally arisen as to whether these calibrations are accurate at high redshift.

With the statistical spectroscopic samples of high-redshift galaxies that now exist, there is evidence that physical properties of high-$z$, star-forming regions are different than what are observed locally. This is typically shown with the well-known offset of the locus of star-forming, high-redshift galaxies relative to that of local, star-forming SDSS galaxies in the [\ion{O}{3}]$\lambda$5007/H$\beta$ vs. [\ion{N}{2}]$\lambda$6583/H$\alpha$ Baldwin$-$Phillips$-$Terlevich \citep[N2-BPT;][]{BPT1981} diagnostic diagram \citep{Steidel2014,Shapley2015,Sanders2016ne_u,Kashino2017,Strom2017}. Numerous studies have tried to explain the primary cause of this evolution with various conclusions. It has been suggested that the offset derives from an elevated ionization parameter \citep{Brinchmann2008,Cullen2016,Kashino2017,Hirschmann2017}, elevated electron density \citep{Shirazi2014}, harder stellar ionizing radiation \citep{Steidel2014,Strom2017,Strom2018}, and/or an increased N/O abundance ratio in high-$z$ galaxies \citep{Masters2014,Shapley2015,Sanders2016ne_u}. It is also possible that there is no single primary cause, and the offset is due to a combination of the aforementioned property evolutions \citep{Kewley2013,Maiolino&Mannucci2019}. Nevertheless, there is considerable motivation to check the validity of locally-calibrated, strong-line metallicity methods at high redshift which utilize the emission lines in the N2-BPT plot and emission lines of other diagnostic diagrams, such as the S2-BPT variant ([\ion{O}{3}]$\lambda$5007/H$\beta$ vs. [\ion{S}{2}]$\lambda\lambda$6716,6731/H$\alpha$) and the $\rm O_{32}$ vs. $\rm R_{23}$ (see Equations \ref{equ:o32} and \ref{equ:r23}, respectively) excitation diagram.

In this paper, we present a detection of the auroral [\ion{O}{3}]$\lambda$4363 emission line in a low-mass, lensed galaxy (\mygal) at $z$ = 2.59. We determine the direct metallicity of \mygal\ and combine it with other (re-calculated) direct metallicity estimates from the literature to examine the applicability of locally-calibrated, oxygen- and hydrogen-based, strong-line metallicity relations at high redshift. In Section \ref{sec:data} of this paper we give an overview of the spectroscopic and photometric observations of \mygal\ and their subsequent reduction. Section \ref{sec:spectrum} discusses the emission-line spectrum of \mygal, highlighting the detection of [\ion{O}{3}]$\lambda$4363 and the method with which the spectrum was fit. Section \ref{sec:properties} examines the physical properties of \mygal\ calculated from the photometry and spectroscopy. Section \ref{sec:discussion} discusses the results of the paper, focusing on the validity and evolution of strong-line metallicity relations with redshift, the evolution of ionization parameter with redshift, the position of \mygal\ in relation to the low-mass end of the FMR, and the position of \mygal\ relative to the predicted MZR from the FIRE hydrodynamical simulations. Section \ref{sec:summary} gives a summary of our results. Appendix \ref{sec:appendix} revisits the [\ion{O}{3}]$\lambda$4363 detection of \citet{Yuan&Kewley2009} with a more sensitive spectrum of the galaxy, taken as part of our larger, dwarf galaxy survey. Throughout this paper, we assume a $\Lambda$CDM cosmology, with $H_0$ = 70 km $\rm s^{-1}$ $\rm Mpc^{-1}$, $\Omega_\Lambda$ = 0.7, and $\Omega_m$ = 0.3.

\begin{figure}[ht!]
    \includegraphics[trim={19.2cm 2cm 18.5cm 1cm}, width=0.48\columnwidth, clip]{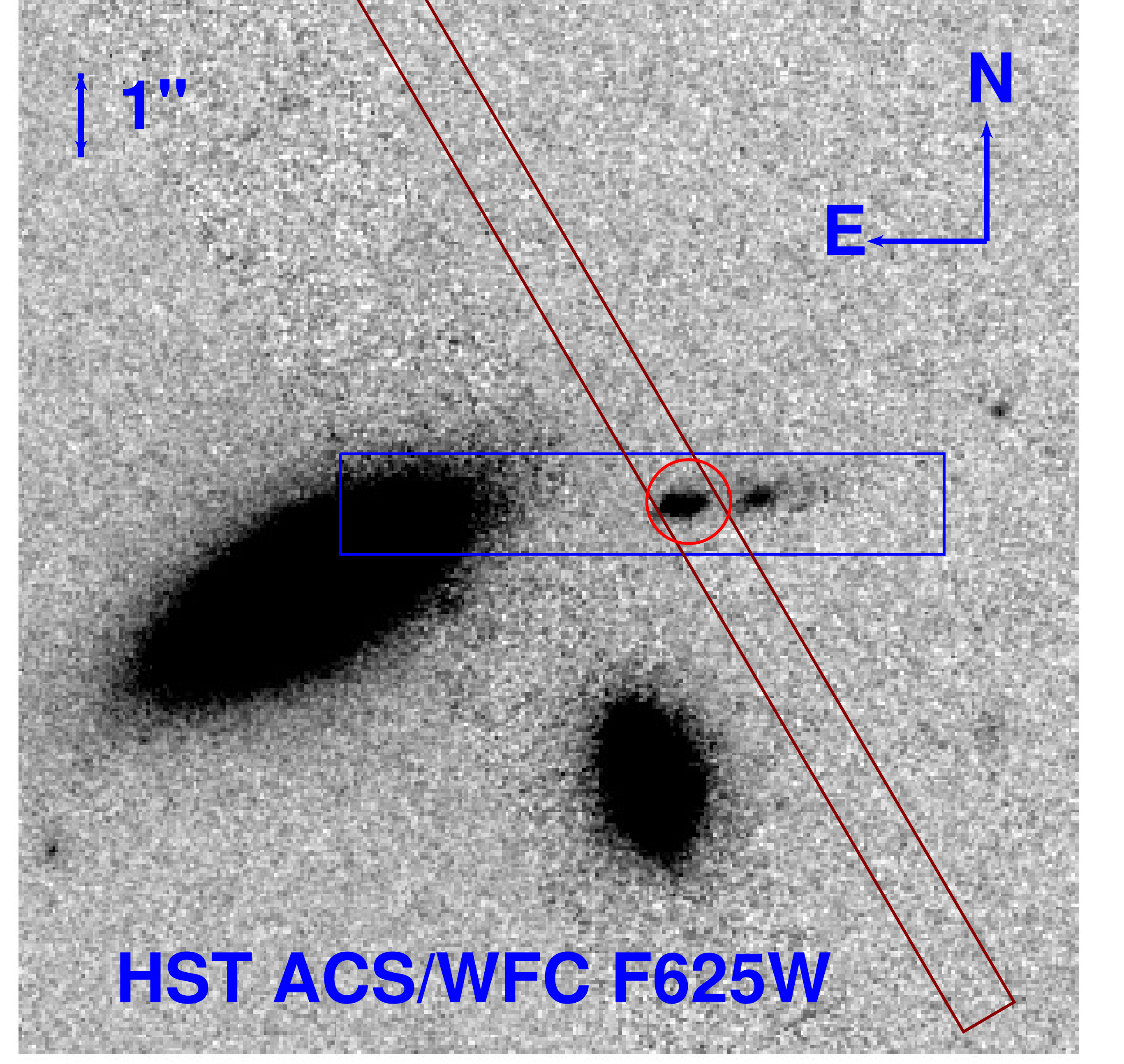}\quad\includegraphics[trim={19.2cm 2cm 18.5cm 1cm}, width=0.48\columnwidth, clip]{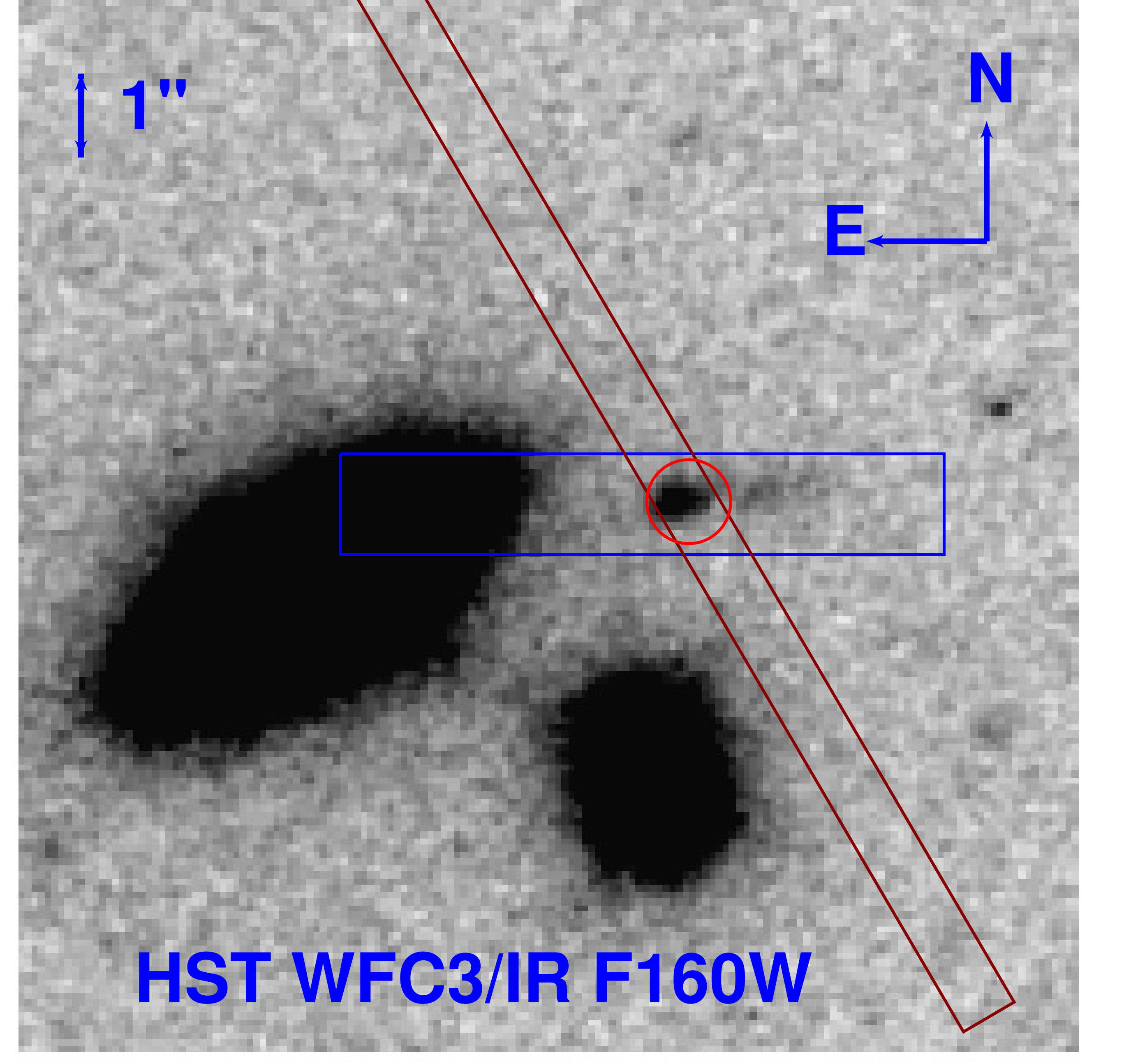}
    \caption{\textit{HST} images of \mygal\ in the ACS/WFC F625W band and WFC3/IR F160W band. The 0\farcs7 MOSFIRE slit is shown in light brown, and the 1\farcs2 LRIS slit is shown in blue. \mygal\ is highlighted by the red circle. Foreground galaxies lie to the south and east of \mygal. Both images are 12\farcs64 on each side.}\label{fig:images}
\end{figure}

\section{Observations and Data Reduction} \label{sec:data}
In this section, we discuss the spectroscopic and photometric observations and reduction for \mygal, lensed by the foreground galaxy cluster Abell 1689. \mygal\ was initially detected via Lyman break dropout selection in the \textit{Hubble Space Telescope} survey of \citet{Alavi2014,Alavi2016}. Based on its photometric redshift and high magnification ($\mu$ = 7.89), it was selected for spectroscopic observation of its rest-frame optical, nebular emission lines as part of a larger spectroscopic survey of star-forming, lensed, dwarf galaxies.

\subsection{Near-IR Spectroscopic Data} \label{subsec:specdata}
Near-IR (rest-optical) spectroscopic data for \mygal\ was taken on 2014 January 2 and 2015 January 17 with the Multi-Object Spectrometer for InfraRed Exploration \citep[MOSFIRE;][]{McLean2010,McLean2012} on the 10-m Keck I telescope. Spectroscopy was taken in the J, H, and K-bands with H-band and K-band data taken the first night (2014) and data in all three bands taken the second night (2015). J-band and H-band data consist of 120 second individual exposures while 180 second exposures were used in the K-band. In total, the integration time is 80 minutes in J-band, 104 minutes in H-band (56 minutes in 2014 and 48 minutes in 2015), and 84 minutes in K-band (60 minutes in 2014 and 24 minutes in 2015). The data were taken with a 0\farcs7 wide slit (see orientation in Figure \ref{fig:images}), giving spectral resolutions of \textit{R} $\sim$ 3310, 3660, and 3620 in the J, H, and K-bands, respectively. An ABBA dither pattern was utilized for all three filters with 1\farcs25 nods for the J-band and 1\farcs2 nods for the H and K-bands.

The spectroscopic data were reduced with the MOSFIRE Data Reduction Pipeline\footnote{\url{ https://keck-datareductionpipelines.github.io/MosfireDRP/}} (DRP). This DRP outputs 2D flat-fielded, wavelength-calibrated, background-subtracted, and rectified spectra combined at each nod position. Night sky lines are used to wavelength-calibrate the J and H-bands while a combination of sky lines and a neon arc lamp is used for the K-band. The 1D spectra were extracted using the IDL software, \texttt{BMEP}\footnote{\url{https://github.com/billfreeman44/bmep}}, from \citet{Freeman2019}. The flux calibration of the spectra was first done with a standard star that was observed at an airmass similar to that of the \mygal\ observations, and then an absolute flux calibration was done using a star included in the observed slit mask.

\subsection{Optical Spectroscopy}
A deep optical (rest-frame UV) spectrum of \mygal\ was taken with the Low Resolution Imaging Spectrometer \citep[LRIS;][]{Oke1995,Steidel2004} on Keck I on 2012 February 24 with an exposure time of 210 minutes. The slit width was 1\farcs2, and the slit was oriented E-W, as seen in Figure \ref{fig:images}. We used the 400 lines/mm grism, blazed at 3400 $\rm \AA$, on the blue side. To reduce read-noise, the pixels were binned by a factor of two in the spectral direction. The resulting resolution is \textit{R} $\sim$ 715. The individual exposures were rectified, cleaned of cosmic rays, and stacked using the pipeline of \citet{Kelson2003}. 

\subsection{Near-UV, Optical, and Near-IR Photometry} \label{subsec:photodata}
Near-UV images of the Abell 1689 cluster, all of them covering \mygal, were taken with the WFC3/UVIS channel on the \textit{Hubble Space Telescope}. We obtained 30 orbits in the F275W filter and 4 orbits in F336W with program ID 12201, followed by 10 orbits in F225W and an additional 14 orbits in F336W (18 orbits total) with program ID 12931. The data were reduced and photometry was measured as described in \citet{Alavi2014,Alavi2016}.

In the optical, we used existing \textit{HST} ACS/WFC images in the F475W, F625W, F775W, and F850LP filters (PID: 9289, PI: H. Ford) as well as in the F814W filter (PID: 11710, PI: J. Blakeslee), calibrated and reduced as detailed in \citet{Alavi2014}. The number of orbits and the 5$\sigma$ depths measured within a 0$\farcs$2 radius aperture for all optical and near-UV filters are given in \citet[][Table 1]{Alavi2016}. In the near-IR, we used existing WFC3/IR images in the F125W and F160W filters (PID: 11802, PI: H. Ford), both with 2,512 second exposure times.

Images of \mygal\ in the optical F625W filter and near-IR F160W filter are shown in Figure \ref{fig:images}.

\begin{figure*}[ht]
    \centering
    \includegraphics[trim={1cm 1.7cm 1.6cm 1.5cm}, width=\textwidth, clip]{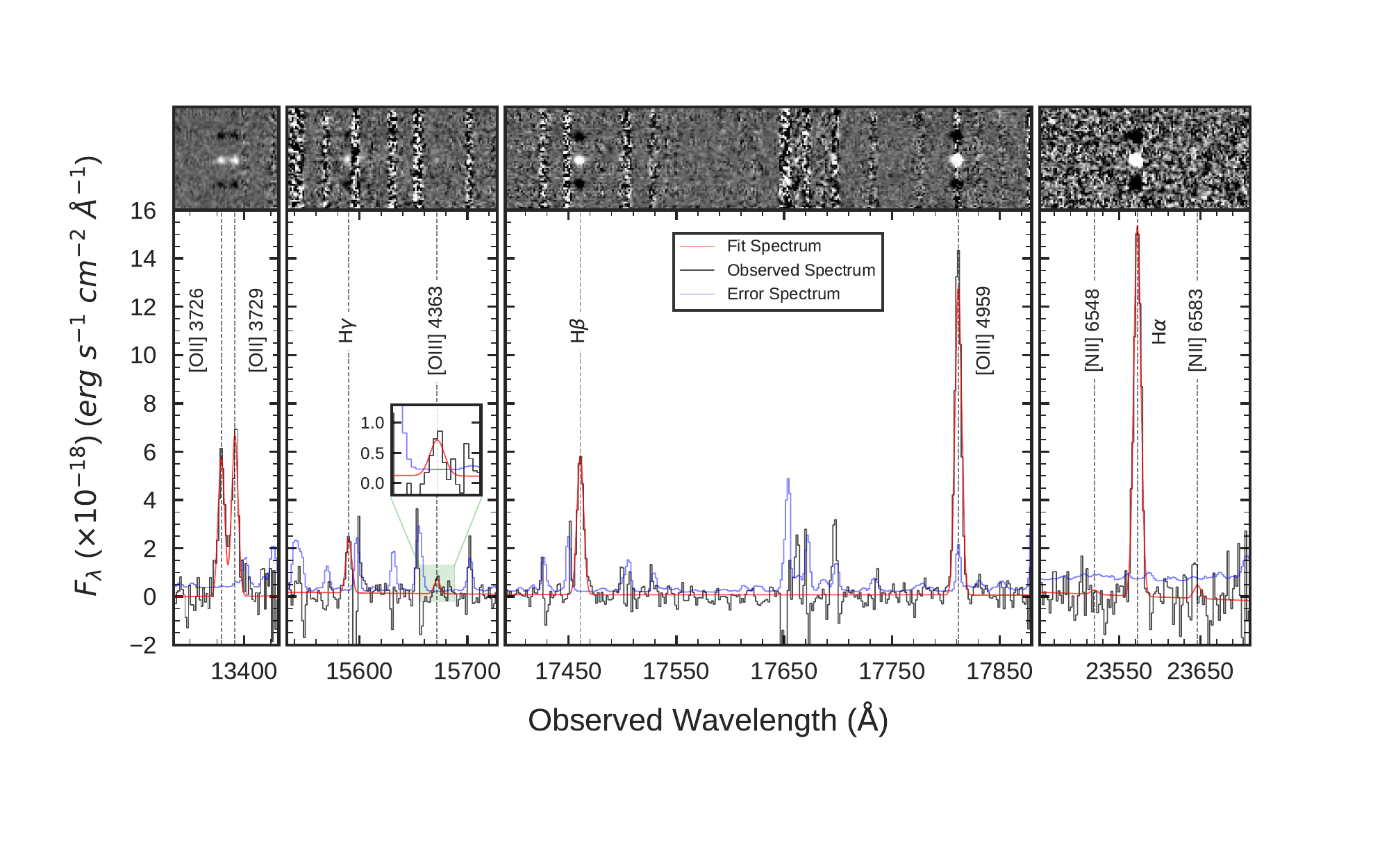}
    \caption{The $z$ = \redshift\ observed spectrum of \mygal\ in the J, H, and K-bands of Keck/MOSFIRE. The top panel shows the two-dimensional spectrum while the bottom panel shows the observed (black), error (blue), and single-Gaussian fit (red) spectra in one dimension. The emission lines are labeled for reference. The portion of the spectrum containing [\ion{O}{3}]$\lambda$4363 has been highlighted in green and magnified in the inset plot. A peak can be seen at the observed location of the line among 4 consecutive pixels with S/N $>$ 1. We report a total significance in the detection of 4.2$\sigma$. Emission of [\ion{O}{3}]$\lambda$4363 in the two-dimensional spectrum is also visible along with the expected symmetric negative images on either side resulting from nodding along the slit.\newline\label{fig:spectrum}}
\end{figure*}

\begin{figure}[t]
    \includegraphics[trim={0.3cm 0.5cm 0.1cm 0cm}, width=\columnwidth, clip]{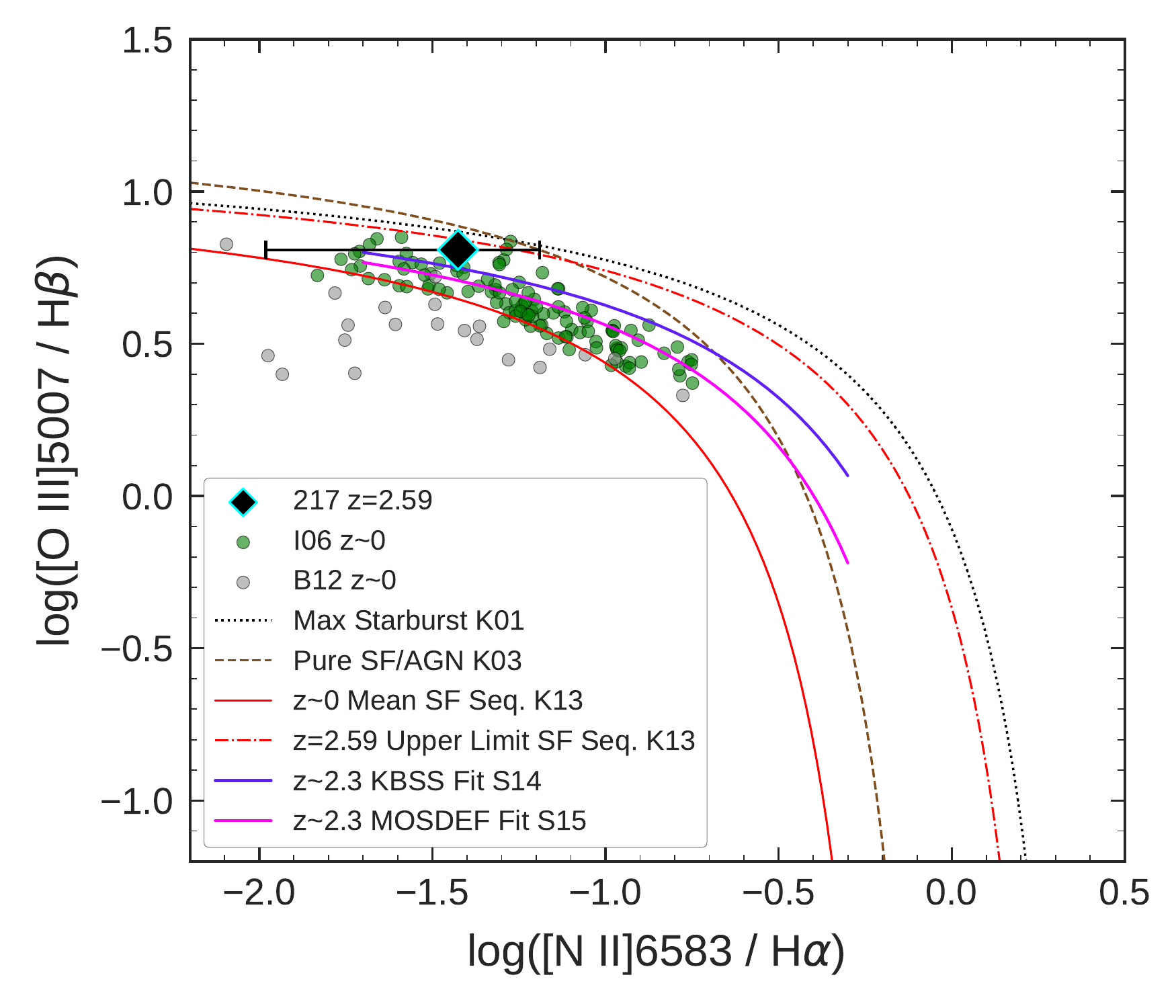}
    \caption{The [\ion{O}{3}]$\lambda$5007/H$\beta$ vs. [\ion{N}{2}]$\lambda$6583/H$\alpha$ N2-BPT diagram. \mygal\ is denoted by the black diamond with cyan border and lies offset from the $z\sim$ 0 mean star-forming sequence of \citet[][K13]{Kewley2013} (solid red line). The galaxy displays high excitation and a very low [\ion{N}{2}]/H$\alpha$ ratio, with the large error bars resulting from the lack of a significant [\ion{N}{2}]$\lambda$6583 detection. The green and gray points represent the $z\sim0$ comparison samples (see Section \ref{sec:discussion}) of \citet[][I06]{Izotov2006} and \citet[][B12]{Berg2012}, respectively. The dotted black line is the ``maximum starburst" curve from \citet[][K01]{Kewley2001}. The dashed brown line is the demarcation between star-forming galaxies and AGN from \citet[][K03]{Kauffmann2003}. The purple line is the best fit to the $z\sim$ 2.3 star-forming galaxies in \citet[][S14]{Steidel2014} while the magenta line is the best fit to the $z\sim$ 2.3 star-forming galaxies in \citet[][S15]{Shapley2015}. The red, dot-dashed line represents the theoretical $z$ = 2.59 upper-limit, star-forming abundance sequence as given by \citet[][K13]{Kewley2013}.\label{fig:bpt}}
\end{figure}

\section{Emission-Line Spectrum of \mygal} \label{sec:spectrum}
\begin{figure}[t]
    \includegraphics[trim={0.2cm 0.55cm 0cm 0cm}, width=\columnwidth, clip]{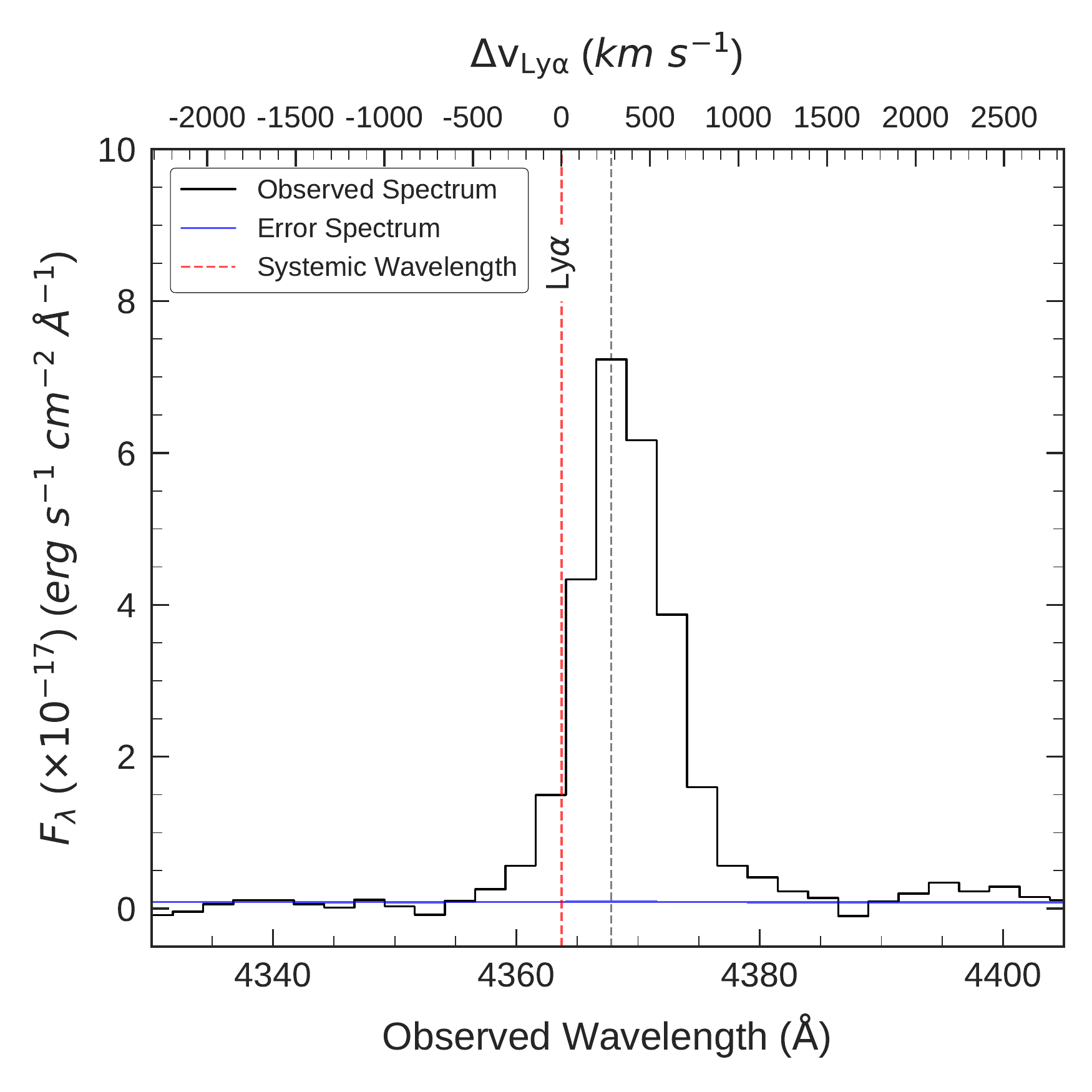}
    \caption{The Ly$\alpha$ emission line of \mygal, observed with Keck/LRIS. The observed and error spectra are shown in black and blue, respectively. The systemic wavelength of Ly$\alpha$ is denoted by the dashed red line. The observed peak of the Ly$\alpha$ line, marked by the dashed gray line, displays a velocity offset (labeled on the upper x-axis) from the systemic redshift of $\Delta v_{\rm\, Ly\alpha}$ = 282 $\rm km\ s^{-1}$.}\label{fig:lya}
\end{figure}

The MOSFIRE spectra yield several emission lines necessary for the direct measurement of intrinsic nebular properties of \mygal, located at $z$ = \redshift\ (see Section \ref{subsec:fitting}). Seen in both 1D and 2D in Figure \ref{fig:spectrum}, we strongly detect [\ion{O}{2}]$\lambda\lambda$3726,3729, H$\gamma$, H$\beta$, [\ion{O}{3}]$\lambda$4959, and H$\alpha$. We also detect the auroral [\ion{O}{3}]$\lambda$4363 line in the H-band (discussed in greater detail in Section \ref{subsec:4363}). The [\ion{O}{3}]$\lambda$5007 emission line, necessary for electron temperature ($T_e$) measurements, is not shown in Figure \ref{fig:spectrum} because it sits at the edge of the H-band filter where transmission declines rapidly, and the flux calibration is uncertain. We instead scale up from the [\ion{O}{3}]$\lambda$4959 line flux using the $T_e$-insensitive intrinsic flux ratio of the doublet, [\ion{O}{3}]$\lambda$5007/[\ion{O}{3}]$\lambda$4959 = 2.98 \citep{Storey&Zeippen2000}. We also note the lack of a significant detection of the [\ion{N}{2}]$\lambda\lambda$6548,6583 doublet in this spectrum, placing \mygal\ in the upper-left corner of the N2-BPT diagnostic diagram as seen in Figure \ref{fig:bpt}. We conclude that \mygal\ is not an AGN based on its very low [\ion{N}{2}]/H$\alpha$ ratio, lack of high-ionization emission lines like [\ion{Ne}{5}], and narrow line widths ($\rm \sigma_{H\beta}$ $\approx$ 53 km $\rm s^{-1}$). The optical spectrum shows strong Ly$\alpha$ emission (see Figure \ref{fig:lya}) with a rest-frame equivalent width of $\rm EW_{0,Ly\alpha}=138$ $\rm \AA$, redshifted by 282 $\rm km\ s^{-1}$. The slit-loss-corrected, observed emission-line fluxes and uncertainties are given in Table \ref{tab:fluxes} with the line-fitting technique described in Section \ref{subsec:fitting}.

\subsection{Detection of [\ion{O}{3}]$\lambda$4363} \label{subsec:4363}
We report a 4.2$\sigma$ detection of the $T_e$-sensitive, auroral [\ion{O}{3}]$\lambda$4363 line. In Figure \ref{fig:spectrum}, there is visible emission in the 2D spectrum at the observed wavelength and spatial coordinates expected for the emission line (as well as the expected symmetric negative images on either side resulting from nodding along the slit). In the magnified inset plot of the highlighted region of the 1D spectrum, there is a clear peak centered at the observed wavelength expected for [\ion{O}{3}]$\lambda$4363 at $z$ = \redshift. We note that this peak is part of 4 consecutive pixels that have a S/N $>$ 1. We also note that at \mygal's redshift, the [\ion{O}{3}]$\lambda$4363 line is not subject to sky line contamination and thus conclude that this detection is robust.

\subsection{Fitting the Spectrum} \label{subsec:fitting}
The spectrum of \mygal\ was fit using the Markov Chain Monte Carlo (MCMC) Ensemble sampler \texttt{emcee}\footnote{\url{https://emcee.readthedocs.io/en/v2.2.1/}} \citep{Foreman-Mackey2013}. In each filter we fit single-Gaussian profiles to the emission lines and a line to the continuum. In the H-band, due to the large wavelength separation between H$\beta$ and [\ion{O}{3}]$\lambda$4363, H$\beta$ and [\ion{O}{3}]$\lambda$4959 were fit separately from H$\gamma$ and [\ion{O}{3}]$\lambda$4363. While the width and redshift were free parameters in the H and K-bands, in the H-band they were only fit with the much higher S/N lines of H$\beta$ and [\ion{O}{3}]$\lambda$4959 and then adopted for H$\gamma$ and [\ion{O}{3}]$\lambda$4363. In the J-band, due to the small wavelength separation of the [\ion{O}{2}] doublet, and thus the partial blending of the lines (seen in Figure \ref{fig:spectrum}), the redshift and width were taken to be the values fit to the highest S/N line in the spectrum (H$\beta$). The redshift of \mygal\ reported in this paper (see Table \ref{tab:properties}) is the weighted average of the redshifts fit to the H and K-bands. 

\begin{figure}[t]
    \includegraphics[trim={0.4cm 0.5cm 0.3cm 0.1cm}, width=\columnwidth, clip]{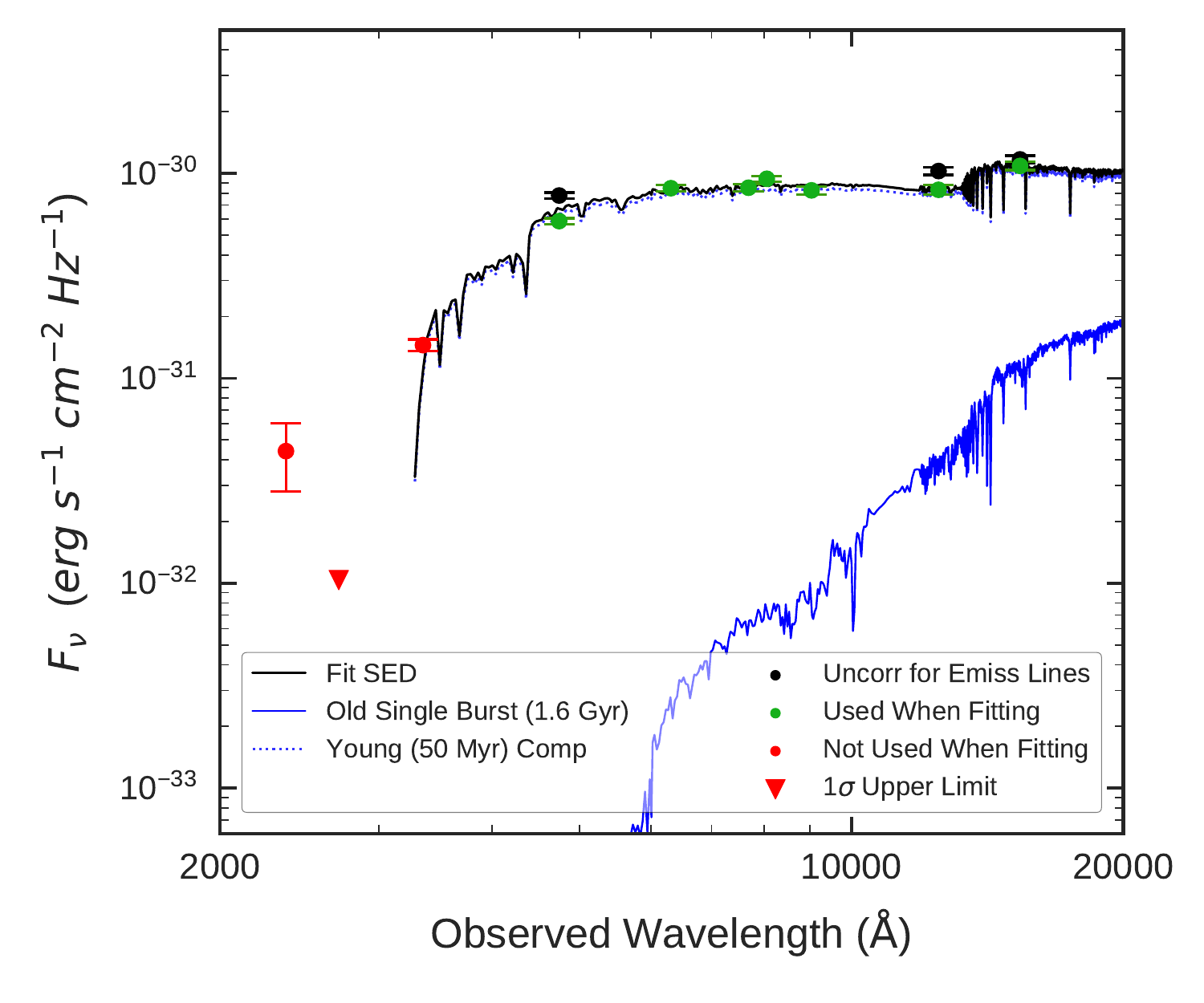}
    \caption{The de-magnified, observed photometry and best-fit SED model (black line) for \mygal. The green data points represent the emission-line-subtracted photometry used for the SED fitting. The black data points represent the photometry before correction for emission lines. The red points signify WFC3/UVIS photometry not used in the fitting because of Ly$\alpha$-forest absorption. An additional 3$\%$ flux error, used to account for systematic errors in the photometry, has been added in quadrature to the flux errors in each of the bands prior to SED fitting and is reflected in the error bars of all (green, black, and red) photometric data points. The SED redshift is fixed to the spectroscopic value of $z_{\rm spec} = \redshift$. The best-fit model indicates a young stellar population ($\sim$ 50 Myrs). Also plotted is a maximally-old (1.6 Gyr) stellar population (blue solid line) that can be added to the fit while slightly scaling down the best-fit, constant-SFR SED (blue dotted line). Adding this older component can increase the stellar mass by a factor of 3.3 at a doubling of the reduced $\chi^2$, so it is treated as an upper limit to the stellar mass.}\label{fig:sed}
\end{figure}

\section{Properties of \mygal}\label{sec:properties}
Estimates of various physical properties of \mygal\ are summarized in Table \ref{tab:properties}, with select properties discussed in greater detail in the sections below.

\begin{deluxetable}{lccrr}[ht!]
\vspace{0.1cm}
\tablecaption{Emission-Line Fluxes and EWs for \mygal\label{tab:fluxes}}
\tablecolumns{5}
\tablenum{1}
\tablewidth{\textwidth}
\setlength{\tabcolsep}{6.8pt}
\tablehead{
\colhead{Line} &
\colhead{$\lambda^\tablenotemark{a}_{\rm rest}$} &
\colhead{$\lambda_{\rm obs}$} &
\colhead{$f^\tablenotemark{b}_{\rm obs}$} &
\colhead{$f^\tablenotemark{b,c}_{\rm corr}$}
}
\startdata
$[$\ion{O}{2}$]$ & 3726.03 & 13$\,$383.21 & 40.8 $\pm$ 1.7 & 222 $\pm$ 9 \\
$[$\ion{O}{2}$]$ & 3728.82 & 13$\,$393.21 & 47.3 $\pm$ 2.2 & 257 $\pm$ 12 \\
H$\gamma$\tablenotemark{d} & 4340.46 & 15$\,$590.12 & 18.3 $\pm$ 1.4 & 81 $\pm$ 6 \\
$[$\ion{O}{3}$]$ & 4363.21 & 15$\,$671.84 & 4.8 $\pm$ 1.1 & 21 $\pm$ 5 \\
H$\beta$\tablenotemark{d} & 4861.32 & 17$\,$460.96 & 53.2 $\pm$ 1.4 & 192 $\pm$ 5 \\
$[$\ion{O}{3}$]$ & 4958.91 & 17$\,$811.48 & 118.7 $\pm$ 4.9 & 414 $\pm$ 17 \\
H$\alpha$\tablenotemark{d} & 6562.79 & 23$\,$572.34 & 206.0 $\pm$ 6.9 & 507 $\pm$ 17 \\
$[$\ion{N}{2}$]$ & 6583.45 & 23$\,$646.52 & 7.8 $\pm$ 5.6 & 19 $\pm$ 14 \\
\hline
\multicolumn{2}{l}{$\rm EW_{0}$(Ly$\alpha$)\tablenotemark{e}} & & \multicolumn{2}{r}{$137.9^{+8.3}_{-8.5}$} \\
\multicolumn{2}{l}{$\rm EW_{0}$([\ion{O}{3}]$\lambda$5007)} & & \multicolumn{2}{r}{860.4 $\pm$ 52.2} \\
\multicolumn{2}{l}{$\rm EW_{0}$(H$\alpha$)} & & \multicolumn{2}{r}{520.7 $\pm$ 28.7} 
\enddata
\tablenotetext{a}{Rest-frame wavelengths in air ($\rm \AA$)}
\tablenotetext{b}{Fluxes are in units of $\rm 10^{-18}\ erg\ s^{-1}\ cm^{-2}$ and are uncorrected for lens magnification. $f_{\rm obs}$ and $f_{\rm corr}$ refer to the observed and dust-corrected fluxes, respectively. Both $f_{\rm obs}$ and $f_{\rm corr}$ are slit-loss-corrected.}
\tablenotetext{c}{The intrinsic flux uncertainties do not include other systematic errors associated with inter-filter calibrations and dust correction, though these additional errors are propagated throughout all of our calculations.}
\tablenotetext{d}{Emission-line fluxes not corrected for underlying stellar absorption as these corrections are small and uncertain (see Section \ref{subsec:ebv_sfr})}
\tablenotetext{e}{Rest-frame equivalent widths in $\rm \AA$}
\tablecomments{The [\ion{O}{3}]$\lambda$5007 line lies at the edge of the H-band filter, so the flux for this line is found via the intrinsic flux ratio of the doublet: [\ion{O}{3}]$\lambda$5007/[\ion{O}{3}]$\lambda$4959 = 2.98}
\vspace{-1.cm}
\end{deluxetable}

\subsection{Stellar Mass and Age}\label{subsec:stellarmass}
The stellar mass is estimated by fitting stellar population synthesis models to the \textit{HST} optical and near-IR photometry. Because some of the emission lines have high equivalent widths (see Table \ref{tab:fluxes}), we have corrected the photometry by subtracting the contribution from the emission lines (e.g., Ly$\alpha$, [\ion{O}{2}]$\lambda\lambda$3726,3729, H$\gamma$, [\ion{O}{3}]$\lambda$4363). We have also added in quadrature an additional 3$\%$ flux error in all bands to account for systematic errors in the photometry \citep{Alavi2016}. We use the stellar population fitting code \texttt{FAST}\footnote{\url{http://w.astro.berkeley.edu/~mariska/FAST.html}} \citep{Kriek2009} with the \citet{BC2003} stellar population synthesis models, and a constant star formation rate with a Chabrier initial mass function \citep[IMF;][]{Chabrier2003}. As suggested by \citet{Reddy2018} for high-redshift, low-mass galaxies, we use the SMC dust extinction curve \citep{Gordon2003} with $A_V$ values varying between $0.0-2.0$. We fix the metallicity at 0.2 Z$_{\odot}$ and the redshift at the spectroscopic value. The stellar age can vary between $7.0<\log(t) \ [$yr$]<10.0$. The $1\sigma$ confidence intervals are derived from a Monte Carlo method of perturbing the broadband photometry within the corresponding photometric uncertainties and refitting the spectral energy distribution (SED) 300 times. The best-fit parameters for \mygal, corrected for the lensing magnification factor, $\mu=7.89$, when necessary, are $A_V=0.25$, $\rm \log(\textit{M}_\ast/\textit{M}_\sun)=8.07$, $\rm SFR=2.75$ M$_{\odot}$ yr$^{-1}$, and $t_{\rm age}\sim 50$ Myr, with the best-fit, de-magnified SED model shown in Figure \ref{fig:sed}. 

The young age of the stellar population is perhaps not surprising as the large H$\alpha$ equivalent width ($\rm EW_{0,H\alpha}=521$ $\rm \AA$) strongly suggests that \mygal\ is undergoing an intense burst of star formation, as seen in a subset of galaxies at high redshift \citep{atek2011,vanderwel2011,straughn2011,atek2014,tang2018}. Because the stellar population associated with this recent burst is young, it has a low mass-to-light ratio and can easily be hiding a significant mass in older stars. To understand how much stellar mass we might be missing, we investigated adding a maximally old stellar population, formed in a single burst at $z=6$ (1.6 Gyr old at $z=\redshift$). We found that the stellar mass could be increased by a factor of 3.3 before the reduced $\chi^2$ is increased by a factor of two (seen in Figure \ref{fig:sed}). Thus, we use $3.3\times$ the mass from the SED fit, or $\rm \log(\textit{M}_\ast/\textit{M}_\sun)<8.59$, as the upper-limit of the stellar mass. 

We note that many of the high-redshift galaxies with [\ion{O}{3}]$\lambda4363$ detections have high equivalent width Balmer lines and may selectively be in a burst relative to the typical galaxy at these redshifts \citep{Ly2015}. Thus, a simple star formation history fit to the photometry might be dominated by the recent burst and will significantly underestimate the stellar mass. This is important to consider when ultimately trying to measure the MZR with these galaxies. 

\subsection{Nebular Extinction and Star Formation Rate}\label{subsec:ebv_sfr}
To properly estimate galactic properties and conditions within the interstellar medium (ISM), several of which rely on flux ratios, the wavelength-dependent extinction from dust must be accounted for. This extinction can be quantified with Balmer line ratios calculated from observed hydrogen emission-line fluxes. With the strong detections of H$\gamma$, H$\beta$, and H$\alpha$ in the spectrum of \mygal, we estimate the extinction due to dust by assuming Case B intrinsic ratios of H$\alpha$/H$\beta$ = 2.79 and H$\alpha$/H$\gamma$ = 5.90 for $T_e$ = 15,000 K and $n_e$ = 100 $\rm cm^{-3}$ \citep{Dopita&Sutherland2003}, approximately the electron temperature and density of \mygal\ (see Section \ref{subsec:edt}).\footnote{The variation in the intrinsic Balmer line ratios with temperature is small over the temperature range typical of \ion{H}{2} regions. We obtain $T_e\sim$ 15,000 K after correcting for dust regardless of using the Balmer ratios corresponding to 15,000 K or the commonly assumed 10,000 K.} We note the presence of underlying stellar absorption of the Balmer lines in Figure \ref{fig:sed} but do not make any corrections to the emission-line fluxes of H$\gamma$, H$\beta$, or H$\alpha$ here as these corrections amount to small percent differences in the fluxes of $\sim3.5\%$, $\sim1.1\%$, and $\sim0.1\%$, respectively, and are also based on an uncertain star formation history. Assuming the extinction curve of \citet{Cardelli1989} with an $R_V$ = 3.1, we find the color excess to be $E(\bv)_{\rm gas} = A_V/R_V =$ 0.39 $\pm$ 0.05. We use this result to correct the observed emission-line fluxes for extinction due to dust and list the corrected values in Table \ref{tab:fluxes}. We note that the nebular extinction is significantly higher than the best-fit extinction of the stellar continuum derived from the SED fit ($A_V=0.25$) and indicated by the flat (in $f_{\nu}$) SED seen in Figure \ref{fig:sed}. This difference in nebular vs. stellar extinction is likely due to the young age of the burst, indicating that the nebular regions are still enshrouded within their birth cloud \citep{charlot&fall2000}. We also note here that some $T_e$-derived metallicities at high redshift are calculated with dust corrections based on the stellar SEDs. If many of these galaxies are in a burst of recent star formation, the stellar attenuation may not be a reliable indicator of the nebular extinction.  This is especially concerning for galaxies with \ion{O}{3}]$\lambda\lambda$1661,1666 detections (rest-UV auroral lines used to estimate $T_e$) instead of [\ion{O}{3}]$\lambda$4363, as the attenuation at these wavelengths is much larger. 

The star formation rate (SFR) of \mygal\ is calculated with the galaxy's dust-corrected H$\alpha$ luminosity ($L$(H$\alpha$)) and the relation between SFR and $L$(H$\alpha$) from \citet{Kennicutt1998}. The conversion factor of the relation is re-calculated assuming a \citet{Chabrier2003} IMF with 0.2 $\rm Z_\sun$, roughly the oxygen abundance of \mygal\ (see Section \ref{subsec:metallicity}). The resulting SFR is divided by the magnification factor ($\mu$ = 7.89) from the lensing model. We estimate that \mygal\ has a SFR = 16.2 $\pm$ 1.8 $\rm M_\sun\ yr^{-1}$. The uncertainty in this measurement does not include the uncertainty in the magnification as the magnification and its error are dependent on the assumptions inherent to the lensing model. We also note here that the H$\alpha$-derived SFR is nearly six times larger than the SED-derived SFR. Much of this discrepancy can be explained if the stellar population has a harder ionizing spectrum due to low Fe abundance \citep{Steidel2014} and/or binary stellar evolution \citep{Eldridge2009}. A harder ionizing spectrum produces more ionizing photons, seen in the H$\alpha$ recombination line, relative to the non-ionizing UV and thus should yield H$\alpha$-based SFRs that are larger than those derived via fitting to rest-UV photometry.

\subsection{Electron Temperature and Density}\label{subsec:edt}
The electron temperature ($T_e$) and electron density ($n_e$) are intrinsic nebular properties that are responsible for the strength of collisionally-excited lines that allow for a direct measurement of the gas-phase metallicity of \ion{H}{2} regions. We calculate the electron temperature in the $\rm O^{++}$ region, $T_e$([\ion{O}{3}]), using the temperature-sensitive line ratio [\ion{O}{3}]$\lambda\lambda$4959,5007/[\ion{O}{3}]$\lambda$4363 and the IRAF task \textsc{nebular.temden} \citep{shaw_dufour1994}. This temperature-sensitive ratio is dependent on electron density, though below $n_e\approx\rm 10^3\ cm^{-3}$ $\sbond$ the low-density regime within which \mygal\ and this paper's literature comparison sample reside $\sbond$ $T_e$([\ion{O}{3}]) is insensitive to the density \citep{Osterbrock_Ferland2006}. We therefore calculate $T_e$([\ion{O}{3}]) non-iteratively, assuming a fiducial electron density of $n_e$ = 150 $\rm cm^{-3}$, appropriate for \ion{H}{2} regions \citep{Sanders2016ne_u}. This yields a result of $T_e$([\ion{O}{3}]) = 14,300 $\pm$ 1,500 K.\footnote{Assuming any $n_e<$ 1,000 $\rm cm^{-3}$ results in variations of our calculated $T_e$ of $<$ 0.5$\%$.} To calculate the electron temperature in the $\rm O^+$ region, $T_e$([\ion{O}{2}]), the auroral doublet [\ion{O}{2}]$\lambda\lambda$7320,7330 is needed. These lines are not within our wavelength coverage, so we utilize the $T_e$([\ion{O}{3}])$\sbond T_e$([\ion{O}{2}]) relation of \citet{Campbell1986} to obtain an electron temperature in the $\rm O^+$ region of $T_e$([\ion{O}{2}]) = 13,000 $\pm$ 1,100 K.

The electron density is estimated with the doublet ratio [\ion{O}{2}]$\lambda$3729/[\ion{O}{2}]$\lambda$3726 and the IRAF task \textsc{nebular.temden}. The aforementioned $T_e$([\ion{O}{2}]) = 13,000 K is used in the calculation. We obtain an electron density for \mygal\ of $n_e$ = $220^{+70}_{-60}$ $\rm cm^{-3}$. This measurement is consistent with the typical electron density found by \citet{Sanders2016ne_u} for $z\sim$ 2.3 star-forming galaxies, $\sim$ 250 $\rm cm^{-3}$, a factor of $\sim$ 10 higher than densities in local star-forming regions. It should be noted, however, that while our measurement agrees with \citet{Sanders2016ne_u} and others \citep[e.g.,][]{Steidel2014,Kashino2017}, our galaxy is $\sim0.9-1.4$ dex lower in stellar mass (see Section \ref{subsec:stellarmass} and Figure \ref{fig:sed}) than the mass ($\rm \sim10^{9.5}\ M_\sun$) above which \citet{Sanders2016ne_u} is confident their density estimate holds true.

\subsection{Oxygen Abundance}\label{subsec:metallicity}
The oxygen abundance, or gas-phase metallicity, is calculated using the analytic ionic abundance expressions of \citet{Izotov2006}. These equations make use of the values found for $T_e$([\ion{O}{2}]), $T_e$([\ion{O}{3}]), and $n_e$ from the previous section. We assume that the oxygen abundance comprises contributions from the populations of the $\rm O^+$ and $\rm O^{++}$ zones of an \ion{H}{2} region with negligible contributions from higher oxygen ionization states.

\begin{equation}\label{equ:o/h}
    \rm \frac{O}{H}\ \approx\ \frac{O^+}{H^+} + \frac{O^{++}}{H^+}
\end{equation}

\noindent We calculate an oxygen abundance for \mygal\ of 12+$\rm \log$(O/H) = 8.06 $\pm$ 0.12 \citep[0.24 $\rm Z_\sun$;][]{Asplund2009}.

\begin{deluxetable}{lr}[ht!]
\tablecaption{Properties of \mygal\label{tab:properties}}
\tablecolumns{2}
\tablenum{2}
\setlength{\tabcolsep}{17pt}
\tablewidth{\textwidth}
\tablehead{
\colhead{Property} &
\colhead{Value}
}
\startdata
R.A. (J2000) & $\rm 13^h\, 11^m\, 27\,\fs\,62$ \\
Dec. (J2000) & $-\rm 01\degr\, 21\arcmin\, 35\,\farcs\,62$ \\
$z$ & 2.591$\,$81 \\
$\;$ & $\pm$ 0.000$\,$01 \\
$\mu$ & 7.89 $\pm$ 0.40 \\
$\rm \log$($M_\ast$/$M_\sun$)\tablenotemark{a,b} & $8.07-8.59$ \\
$M_{\rm UV,1700}$\tablenotemark{a} & $\rm -18.67$ $\pm$ 0.04 \\
$E(\bv)_{\rm gas}$ & 0.39 $\pm$ 0.05 \\
SFR\tablenotemark{a} [$\rm M_\sun$ $\rm yr^{-1}$] & 16.2 $\pm$ 1.8 \\
$n_e$ [$\rm cm^{-3}$] & $220^{+70}_{-60}$ \\
$T_e$([\ion{O}{2}]) [K] & 13$\,$000 $\pm$ 1100 \\
$T_e$([\ion{O}{3}]) [K] & 14$\,$300 $\pm$ 1500 \\
12+$\rm \log$($\rm O^{+}/H^{+}$) & 7.56 $\pm$ 0.12 \\
12+$\rm \log$($\rm O^{++}/H^{+}$) & 7.90 $\pm$ 0.12 \\
12+$\rm \log$(O/H) & 8.06 $\pm$ 0.12 \\
$Z$ [$\rm Z_\sun$] & $0.24^{+0.08}_{-0.06}$
\enddata
\tablenotetext{a}{Most probable value corrected for the listed magnification factor, $\mu$. The uncertainty does not include the uncertainty in the magnification.}
\tablenotetext{b}{The lower and upper bounds of the stellar mass estimate. The lower bound corresponds to our best-fit SED model ($t\sim$ 50 Myr), and the upper bound corresponds to a young stellar component ($t$ = 50 Myr) in combination with a 1.6 Gyr old burst component. See Section \ref{subsec:stellarmass} and Figure \ref{fig:sed} for further details.}
\vspace{-1.1cm}
\end{deluxetable}

\subsection{Uncertainties}\label{subsec:uncertainties}
To calculate the 1$\sigma$ uncertainties of the intrinsic emission-line fluxes, flux ratios, and other properties of \mygal, we utilize a Monte Carlo approach in which a given value is sampled $N$ = $10^5$ times. The uncertainties in the intrinsic emission-line fluxes are found by first sampling the probability distribution of \mygal's extinction in the visual band ($A_V$), needed for the extinction at a given wavelength ($A_\lambda$), and the probability distribution of each emission line's observed flux. The final probability distribution of $A_V$ is the result of multiplying the probability distributions of $A_V$ found for each of the Balmer decrements considered for \mygal, H$\alpha$/H$\beta$ and H$\alpha$/H$\gamma$, the uncertainty for each ratio coming from its observed statistical error added in quadrature with a 5$\%$ inter-filter systematic error. The visual-band extinction and the emission lines are each sampled $N$ times from a normal distribution centered on the most probable $A_V$ or observed flux, respectively, with a standard deviation given by the 1$\sigma$ error of the value being sampled. The $A_V$ values are then used to calculate $N$ extinction magnitudes for each emission line, with which each iteration of each emission-line sample is dust-corrected, giving a sample of $N$ intrinsic fluxes for each line. A posterior histogram is then generated for the intrinsic flux of each line, and a 68$\%$ confidence interval is fit, allowing a 1$\sigma$ uncertainty to be determined for each line's intrinsic flux.

In the calculation of the flux-ratio uncertainties, we take the samples of intrinsic emission-line fluxes and calculate $N$-length samples of the desired flux ratios, for which posterior histograms are created and 1$\sigma$ errors estimated as for the intrinsic emission-line fluxes. The properties of \mygal\ have their uncertainties estimated in the same manner. 

\section{Discussion}\label{sec:discussion}
\subsection{Strong-Line Ratio $-$ Metallicity Diagnostics}\label{subsec:strong-lines}
Having calculated the intrinsic emission-line fluxes and direct-metallicity estimate of \mygal, we study the evolution of both nebular physical properties and the relationships between strong-line ratios and $T_e$-based metallicities. 

\citet{Jones2015} presented the first calibrations between strong-line ratios and direct metallicities at significant redshift, utilizing a sample of 32 star-forming galaxies at $z\sim$ 0.8 from the DEEP2 Galaxy Redshift Survey \citep{Davis2003,Newman2013}. Because the flux ratio of [\ion{O}{3}]$\lambda$4363/[\ion{O}{3}]$\lambda$5007 is generally $\lesssim 3\%$, random noise creates a large scatter in the measurement of this temperature-sensitive ratio. To combat this effect, all 32 galaxies in the Jones et al. sample were selected because they have high S/N in [\ion{O}{3}]$\lambda$5007 and low noise in the location of [\ion{O}{3}]$\lambda$4363. More specifically, the galaxies in the sample have a ratio of [\ion{O}{3}]$\lambda$5007 flux to uncertainty in the [\ion{O}{3}]$\lambda$4363 flux ($f_{5007}/\sigma_{4363}$) of $\geqslant$ 300. This ratio, which they call the ``sensitivity" (this term used hereafter to denote this ratio), not only reduces the effects of random noise but also the bias toward very low metallicity (12+$\rm \log$(O/H) $\lesssim$ $8.3-8.4$ or $Z\lesssim$ $0.4-0.5$ $\rm Z_\sun$) galaxies that comes with selecting a sample via [\ion{O}{3}]$\lambda$4363 significance instead (see their Figure 1). 

\citet{Jones2015} found that the relations between direct metallicity and ratios of neon, oxygen, and hydrogen emission lines derived from their sample are consistent (albeit with larger uncertainties) with the relations derived from a subset (subject to the same sensitivity requirement) of the $z\sim$ 0 star-forming galaxies from \citet{Izotov2006} $\sbond$ a subsample itself from Data Release 3 of the Sloan Digital Sky Survey \citep{Abazajian2005}. Jones et al. showed that these relations do not evolve from $z$ = 0 to $z\sim$ 0.8. 

\subsubsection{Comparison Samples Across Cosmic Time}\label{subsubsec:comparison}
In a similar manner to \citet{Jones2015} and \citet{Sanders2016o3} with their object COSMOS-1908, we will use the measurements of \mygal, compared to other [\ion{O}{3}]$\lambda$4363 sources at various redshifts, to further study the evolution of the calibrations in \citet{Jones2015}, particularly at higher redshift. We note that unlike in \citet{Jones2015} and \citet{Sanders2016o3}, the relations involving [\ion{Ne}{3}]$\lambda$3869 are not studied here because this line falls out of our spectroscopic coverage of \mygal. 

In addition to the 32, $z\sim$ 0.8 galaxies from \citet{Jones2015}, we also consider two local, $z\sim$ 0 comparison samples:  113 star-forming galaxies with spectral coverage of the optical [\ion{O}{2}] doublet from \citet{Izotov2006} $\sbond$ the same $z\sim$ 0 sample used in \citet{Jones2015} $\sbond$ and 28 \ion{H}{2} regions (21 total galaxies) from \citet{Berg2012}. The galaxies from \citet{Berg2012} comprise a low-luminosity subsample of the \textit{Spitzer} Local Volume Legacy (LVL) catalog \citep{Dale2009} and have high-resolution MMT spectroscopy for [\ion{O}{3}]$\lambda$4363 detection. This particular sample was chosen because of its low-luminosity and the volume-limited $\sbond$ as opposed to flux-limited $\sbond$ nature of its parent LVL sample, the combination of which allows for the statistical study of local dwarf galaxies ($\rm 5.90\ \leqslant\ \log$($M_\ast/M_\sun$)$\ \leqslant\ \rm 9.43$ here). These Berg et al. sample qualities are similar to those of our high-$z$ parent survey, to which \mygal\ belongs, in the sense that we are looking at very low-mass objects (via lensing) in a small volume as opposed to less-typical, more luminous objects in a larger volume. 

Both of the local comparison samples adhere to the sensitivity cut placed on the \citet{Jones2015} sample. Additionally, as in \citet{Izotov2006}, we arrived at our stated comparison sample sizes by removing all galaxies (or \ion{H}{2} regions) with both [\ion{O}{3}]$\lambda$4959/H$\beta$ $<$ 0.7 and [\ion{O}{2}]$\lambda$3727/H$\beta$ $>$ 1.0, ensuring high-excitation samples that do not discriminate against very metal-deficient sources with high excitation. Global oxygen abundance and strong-line ratio values for galaxies in the \citet{Berg2012} sample with multiple \ion{H}{2} regions meeting these cuts are taken as the average of the individual \ion{H}{2} region values, weighted by the uncertainties calculated for the abundances and ratios, respectively, as detailed in Section \ref{subsec:uncertainties}.

At low-to-intermediate redshifts, we also include 9 of the 20, $z<0.9$, high-sSFR galaxies with [\ion{O}{3}]$\lambda$4363 detections from \citet{Ly2014} and the Subaru Deep Field \citep{Kashikawa2004}, excluding the rest of the sample due to the inability to determine dust corrections, unreliable $T_e$ estimates, missing H$\beta$ or stellar mass (necessary for our study of the FMR in Section \ref{subsec:fmr}), and the presence of a LINER. Due to this sample being so small, we do not apply the sensitivity cut of \citet{Jones2015}, which would remove 5 of the 9 objects, but note that all galaxies pass the cut of \citet{Izotov2006}.

\begin{figure*}[ht!]
    \centering
    \includegraphics[trim={0.25cm 0.25cm 0cm 0cm}, width=\textwidth, clip]{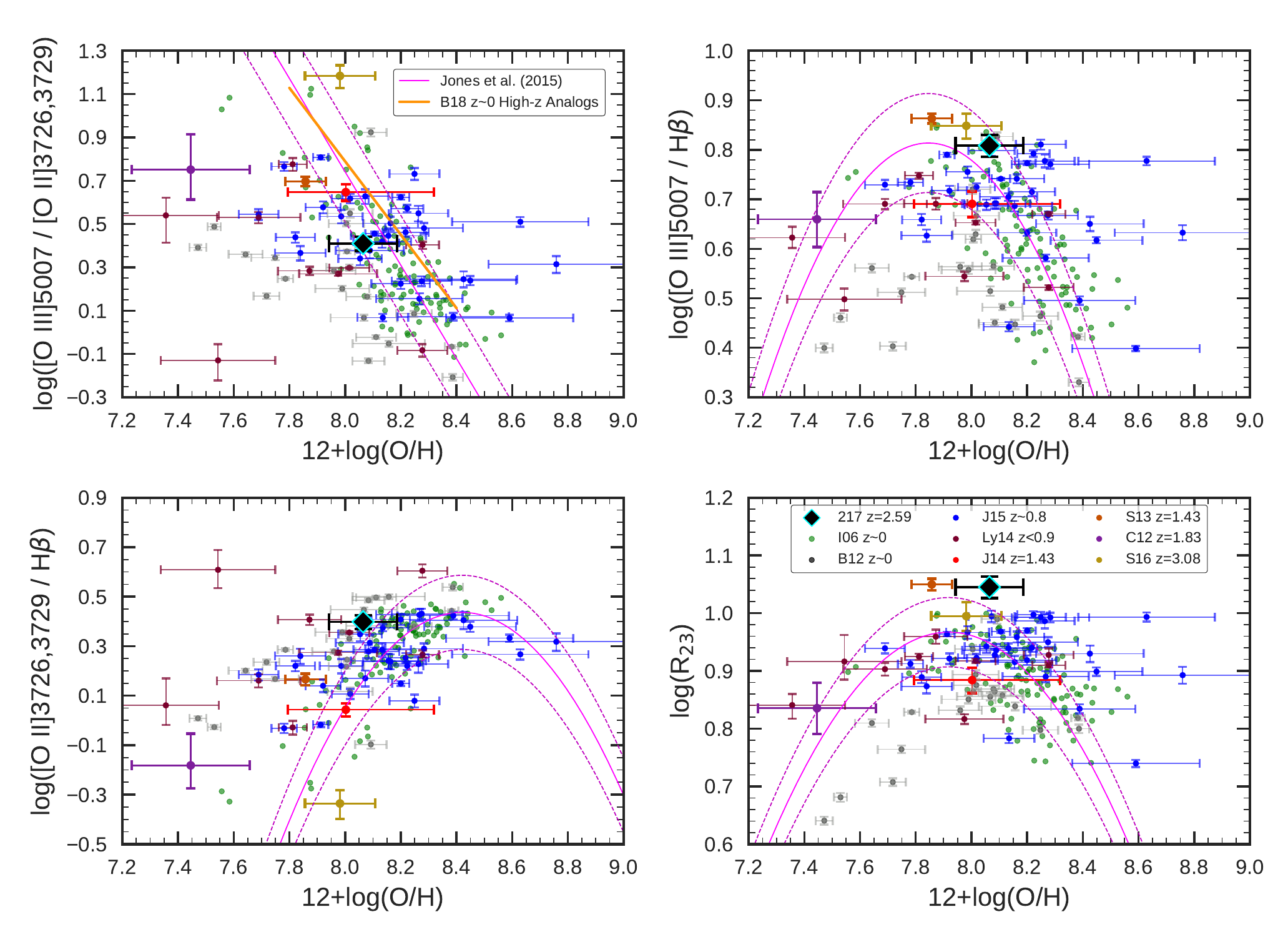}
    \caption{Strong emission-line ratios vs. direct-method oxygen abundance for \mygal\ and comparison samples ranging in redshift from $z\sim$ 0 to $z\sim$ 3.1. \mygal\ is denoted by the black diamond with cyan border. The $z\sim$ 0 sample of \citet[][I06]{Izotov2006} is given by the green points. The $z\sim$ 0.8 sample of \citet[][J15]{Jones2015} is given by the blue data points. The $z<0.9$ sample of \citet[][Ly14]{Ly2014} is given by the dark red data points. The red, dark orange, purple, and gold points correspond to the $z$ = 1.43 galaxy of \citet[][J14]{James2014}, the $z$ = 1.43 galaxy of \citet[][S13]{Stark2013}, the $z$ = 1.83 galaxy of \citet[][C12]{Christensen2012}, and the $z$ = 3.08 galaxy of \citet[][S16]{Sanders2016o3}, respectively. The solid magenta lines show the best-fit relations between the strong-line ratios and metallicity as determined by \citet{Jones2015} with the $z\sim$ 0 \citet{Izotov2006} sample. The accompanying dashed magenta lines represent the 1$\sigma$ intrinsic scatter around the best-fit relations. The orange line in the upper-left panel is the best-fit relation, based on stacked spectra of $z\sim$ 0 high-$z$ analogs, of \citet[][B18]{Bian2018}. The sample of \citet[][B12]{Berg2012} $z\sim$ 0 LVL galaxies is represented by the gray points and included to show the disparity between this low-excitation (see also Figure \ref{fig:bpt}), low-sSFR (median sSFR $\sim$ 0.2 $\rm Gyr^{-1}$ for the objects used here) sample and the other comparison samples when investigating these strong-line ratio $\sbond$ metallicity relations.\newline\label{fig:strong-lines}}
\end{figure*}

In addition to the low- and intermediate-redshift samples, we also compare \mygal\ to the galaxies of \citet{James2014} at $z$ = 1.43, \citet{Stark2013} at $z$ = 1.43, \citet{Christensen2012} at $z$ = 1.83, and \citet{Sanders2016o3} at $z$ = 3.08. Each of these galaxies has an [\ion{O}{3}]$\lambda$4363 detection and corresponding, re-calculated, direct metallicity estimate. We do not compare to the galaxy reported in \citet{Yuan&Kewley2009} as our deeper spectrum of this galaxy shows that the claimed [\ion{O}{3}]$\lambda$4363 detection is not correct. See Appendix \ref{sec:appendix} for more details. All comparison samples in this paper, at $z\sim$ 0$-$3.1, are dust-corrected using the \citet{Cardelli1989} extinction curve, with an $R_V$ = 3.1 (except for \citealt{Jones2015}, who use an $R_V$ = 4.05 though show that their results are insensitive to this value), and have had their physical properties re-calculated using the methods detailed in Sections \ref{subsec:edt} and \ref{subsec:metallicity}.

We do not include any \ion{O}{3}]$\lambda\lambda$1661,1666 sources in our comparison samples as do some other similar studies \citep[e.g.,][]{Patricio2018,Sanders2019} due to added complications when considering both the optical and ultraviolet. These complications lie primarily in the very uncertain extinction law in the UV and the large wavelength separation between these auroral lines and [\ion{O}{3}]$\lambda$5007, as well as in issues arising from observing in these different regimes (e.g., different instruments, slit widths, seeing).

\subsubsection{The Evolution of the Strong-Line Ratio $-$ Metallicity Calibrations}\label{subsubsec:slr-interp}
In our effort to further quantify the evolution at high redshift of the locally-calibrated, strong-line metallicity relations, as well as other physical properties, we consider the position of \mygal, and the other high-redshift galaxies, in relation to the \citet{Jones2015} calibrations and other lower-redshift comparison samples in the four panels of Figure \ref{fig:strong-lines}. We find that \mygal\ is consistent with the local best-fit relations of \citet{Jones2015} in the top two and bottom-left panels, given \mygal's uncertainties and the relations' intrinsic scatter. We observe \mygal\ to be $\sim$1.6$\sigma$ above the best-fit $\rm R_{23}$ (see Equation \ref{equ:r23} for $\rm R_{23}$ ratio) relation at its metallicity of $Z$ = 8.06, though we do not claim it to be inconsistent with the relation based on \mygal's uncertainties in both parameters, especially oxygen abundance, combined with the scatter around the relation. \mygal's elevated $\rm R_{23}$ value is a consequence of \mygal\ being above the local relation in the [\ion{O}{3}]$\lambda$5007/H$\beta$ ratio and especially in the [\ion{O}{2}]$\lambda\lambda$3726,3729/H$\beta$ ratio, though both ratios are consistent with the local calibrations. When also considering the other $z >$ 1 sources in addition to \mygal, we do not observe any significant systematic offsets in line ratio or metallicity for any of the relations. We therefore suggest that there is no evidence of evolution from $z\sim$ 0 to $z\sim$ 3.1 in the relations between direct metallicity and emission-line ratios involving only oxygen and hydrogen. However, larger samples of [\ion{O}{3}]$\lambda$4363 detections are needed in order to significantly constrain the evolution out to high redshift. 

We do caution, however, that 4 out of the 5 $z >$ 1 galaxies lie at or very near the turnover portion of the [\ion{O}{3}]$\lambda$5007/H$\beta$ and $\rm R_{23}$ relations, where variation in the strong-line ratio is small over the corresponding oxygen abundance range, limiting the constraining power of the relations when determining the metallicity at fixed line-ratio. This is seen as well in the recent work of \citet{Sanders2019}, who study the relationships between strong-line ratios and direct metallicity using a sample of 18 galaxies at 1.4 $\lesssim\ z\ \lesssim$ 3.6 with [\ion{O}{3}]$\lambda$4363 or \ion{O}{3}]$\lambda\lambda$1661,1666 auroral-line detections, including 3 new [\ion{O}{3}]$\lambda$4363 detections from the MOSFIRE Deep Evolution Field survey \citep[MOSDEF;][]{Kriek2015}. They show an abundance of objects with 7.7 $<$ 12+$\log$(O/H) $<$ 8.1 lying at these turnovers and caution against the use of these line ratios at high-$z$ for galaxies within this metallicity regime.

In addition to the strong-line metallicity relations of \citet{Jones2015}, we plot the [\ion{O}{3}]/[\ion{O}{2}] $\sbond$ direct metallicity calibration of \citet{Bian2018} (top-left panel of Figure \ref{fig:strong-lines}), who utilized stacked spectra with [\ion{O}{3}]$\lambda$4363 of $z\sim$ 0 high-$z$ analogs that lie at the same location on the N2-BPT diagram as $z\sim$ 2.3 star-forming galaxies. This calibration is favored in \citet{Sanders2019} for its linear relation between the strong-line ratio and metallicity, its ability to closely reproduce ($\sim$ 0.1 dex) the average metallicity of their $z>$ 1 sample, and its derivation from an analog sample selected via strong-line ratios rather than global galaxy properties. Within the range of applicability, $12+\log\rm{(O/H)} = 7.8 - 8.4$, there is generally good agreement between the relation, our various samples (including \mygal), and the relation of \citet{Jones2015} as the relation of \citet{Bian2018} lies within the intrinsic scatter around that of \citet{Jones2015}.

We note that the majority of the \citet{Berg2012} line ratios do not follow the local relations with direct metallicity. While there is good agreement between the local Jones et al. relations and the few \ion{H}{2} regions in the Berg et al. sample with $8.2 \lesssim 12+\log\rm{(O/H)} \lesssim 8.4$, the bulk of the \ion{H}{2} region sample, having $12+\log\rm{(O/H)} \lesssim 8.1$, lies removed from these relations. This is seen as well in the strong-line ratio $\sbond$ direct metallicity plots of \citet[][Figure 3]{Sanders2019}, who find agreement at $12+\log\rm{(O/H)}\sim8.3$ between the median relations of individual $z=0$ \ion{H}{2} regions and their $z\sim0$ and $z >$ 1 galaxy samples, but similar divergences below an oxygen abundance of $\sim$ 8.0. As Sanders et al. suggests, this may be due to an incomplete sample of local, high-excitation, low-metallicity \ion{H}{2} regions, possibly a result of the short-lived nature of individual star-forming regions and their rapidly changing ionizing spectra.

\subsection{$\rm O_{32}$ vs. $\rm R_{23}$ Excitation Diagram and its Use as a Metallicity Indicator}\label{subsec:o32_r23}
The $\rm O_{32}$ vs. $\rm R_{23}$ excitation diagram relates optical emission-line ratios given by the following equations:

\begin{equation}\label{equ:o32}
    \rm O_{32}\ =\ \frac{[\text{\ion{O}{3}}]\lambda\lambda4959,5007}{[\text{\ion{O}{2}}]\lambda\lambda3726,3729}
\end{equation}

\begin{equation}\label{equ:r23}
   \rm R_{23}\ =\ \frac{[\text{\ion{O}{2}}]\lambda\lambda3726,3729+[\text{\ion{O}{3}}]\lambda\lambda4959,5007}{\rm H\beta}
\end{equation}

\begin{figure*}[t!]
    \includegraphics[trim={0cm 0.3cm 0cm 0cm},width=\columnwidth,height=7.62cm,clip]{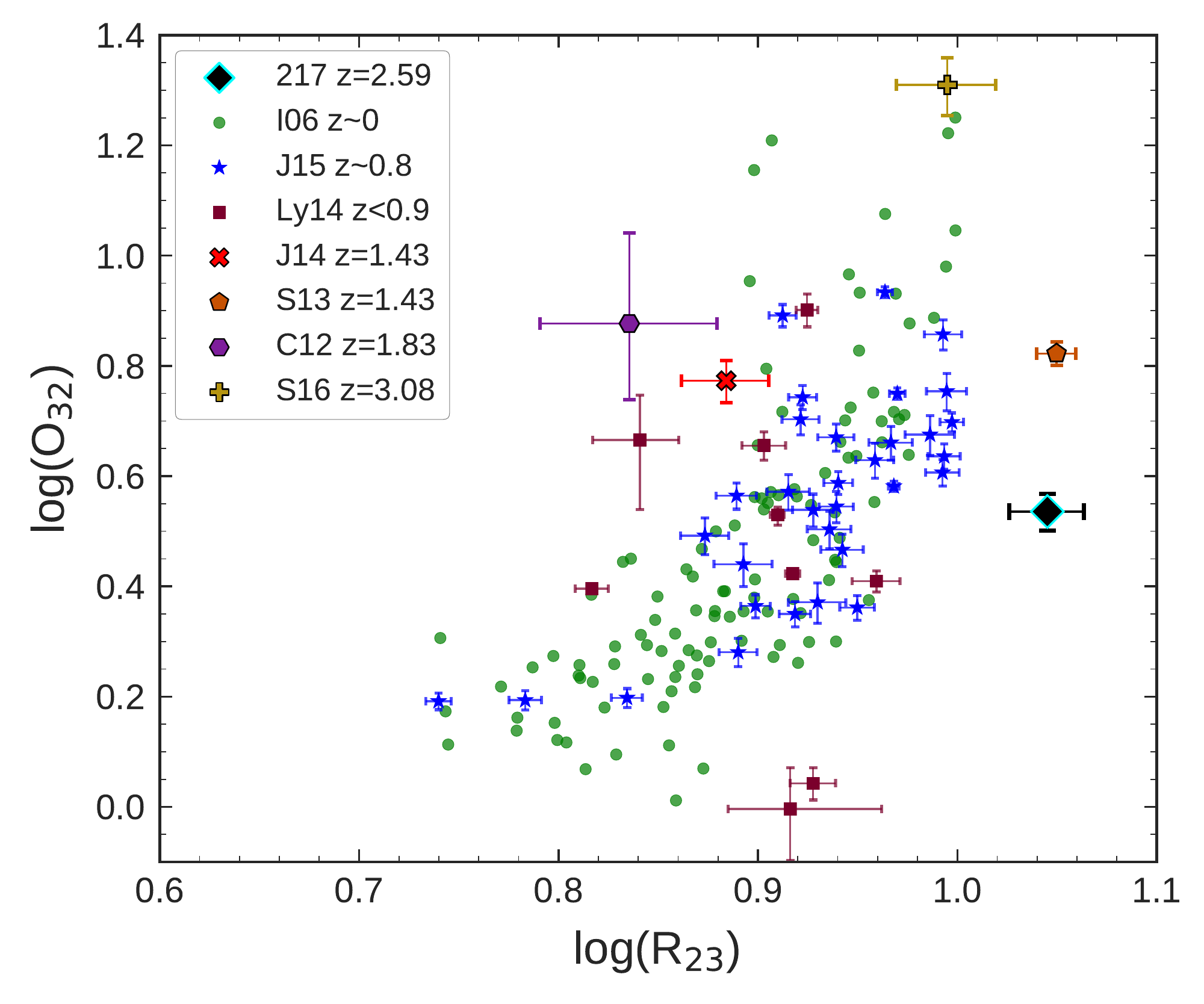}\quad\includegraphics[trim={0cm 0.3cm 0cm 0cm},width=1.09\columnwidth,clip]{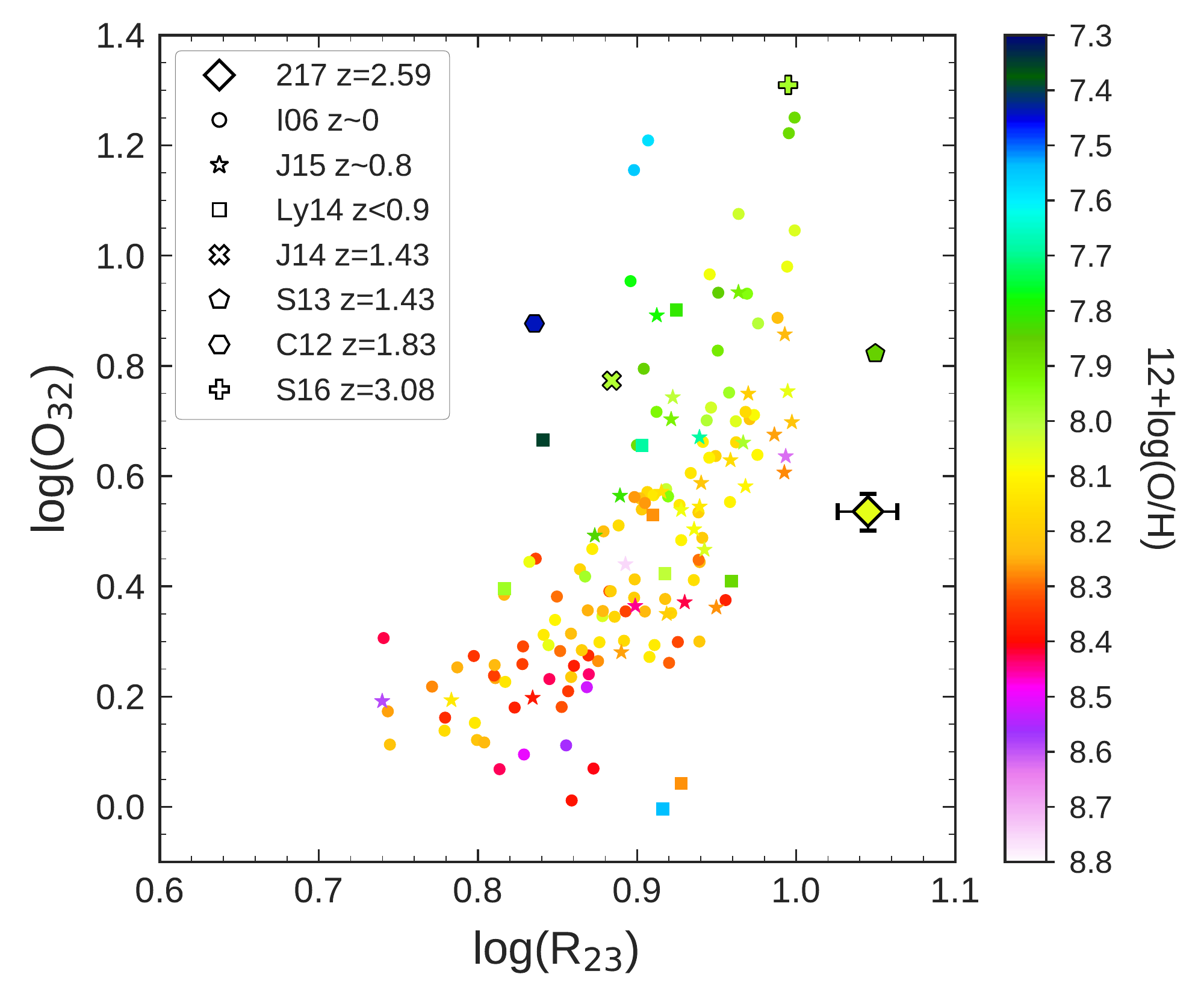}
    \caption{High-excitation tail of the $\rm O_{32}$ vs. $\rm R_{23}$ excitation diagram. (Left) \mygal\ and the comparison samples, with error bars, following the same color scheme as in Figure \ref{fig:strong-lines}. \mygal\ is represented by a diamond with a cyan border. (Right) \mygal\ and the comparison samples color-coded by their direct metallicity estimates. \mygal\ is again represented by a diamond, now with a black border. In both panels, the \citet[][I06]{Izotov2006} sample is represented by circles. The \citet[][J15]{Jones2015} sample is represented by stars. The \citet[][Ly14]{Ly2014} sample is denoted by squares. The \citet[][J14]{James2014} galaxy, \citet[][S13]{Stark2013} galaxy, \citet[][C12]{Christensen2012} galaxy, and \citet[][S16]{Sanders2016o3} galaxy are given by an $\times$, pentagon, hexagon, and plus sign, respectively. The color mapping of this plot demonstrates the roughly monotonic and redshift-independent decrease in oxygen abundance from low-to-high $\rm O_{32}$ and $\rm R_{23}$ as first demonstrated empirically by \citet{Shapley2015}.\newline\label{fig:o32r23}}
\end{figure*}

\noindent As seen in the high-excitation tail of $\rm O_{32}$ vs. $\rm R_{23}$ displayed in Figure \ref{fig:o32r23} for \mygal\ and the comparison samples, as well as in full in the literature \citep[e.g.,][]{Nakajima2013,Nakajima&Ouchi2014,Shapley2015,Sanders2016ne_u,Strom2017}, the excitation diagram characteristically has a strong correlation between higher $\rm O_{32}$ and $\rm R_{23}$ values. It has also been shown by \citet{Nakajima&Ouchi2014} with a sample of $z=2-3$ Lyman Break Galaxies (LBGs), \citet{Shapley2015,Sanders2016ne_u} with $z\sim$ 2.3 galaxies from the MOSDEF survey, and \citet{Strom2017} with $z\sim$ 2.3 galaxies from the KBSS survey that high-redshift, star-forming galaxies follow the same distribution as local SDSS galaxies toward higher $\rm O_{32}$ and $\rm R_{23}$ values. Indeed, when looking at the galaxies in the left panel of Figure \ref{fig:o32r23}, we see no evidence for significant evolution at any of the redshifts considered by our samples. 

Individually, the $\rm O_{32}$ ratio serves as a commonly used diagnostic of the ionization parameter of a star-forming region \citep[see][]{Kewley&Dopita2002,Sanders2016ne_u} while the $\rm R_{23}$ ratio is a commonly used diagnostic for the gas-phase oxygen abundance of a star-forming region \citep{Pagel1979}. However, as detailed in \citet{Kewley&Dopita2002}, $\rm O_{32}$ is dependent on metallicity, and $\rm R_{23}$ is dependent on the ionization parameter. Furthermore, as seen in Figure \ref{fig:strong-lines}, the $\rm R_{23}$ diagnostic is double-valued \citep{Kewley&Dopita2002} and not very sensitive to the majority of the sub-solar oxygen abundances studied in this work. The variation of $\sim$ 0.3 dex in $\log(\rm R_{23})$ seen here in Figures \ref{fig:strong-lines} and \ref{fig:o32r23} supports the findings of \citet[][see Figure 11]{Steidel2014}, who show, via photoionization models, that $\log(\rm R_{23})$ is nearly independent of input oxygen abundance in high-redshift galaxies with gas-phase metallicities ranging from 0.2$-$1.0 $\rm Z_\sun$. 

If instead these two ratios are considered simultaneously in the $\rm O_{32}$ vs. $\rm R_{23}$ excitation diagram, the double-valued nature of the $\rm R_{23}$ diagnostic is removed, and a combination of ionization parameter and metallicity can be obtained. \citet{Kewley&Dopita2002}, \citet{Nakajima2013}, \citet{Nakajima&Ouchi2014}, and \citet{Strom2018} have all utilized this excitation diagram in combination with photoionization models to calculate oxygen abundances, out to $z\sim$ 2 in the latter three studies. \citet{Shapley2015} took an empirical approach to suggesting this excitation diagram's value as an abundance indicator, using the direct metallicity estimates from stacked SDSS spectra of \citet{Andrews&Martini2013} to show a nearly monotonic decrease in metallicity from low-to-high $\rm O_{32}$ and $\rm R_{23}$. They showed that while $\rm R_{23}$ considered alone does not vary greatly with metallicity, the position within the 2D space defined by these two line ratios correlates strongly with metallicity. They further argued that due to the apparent lack of evolution in high-redshift galaxies along the high-excitation end of the diagram, a redshift-independent (out to $z\sim$ 2.3, at least) metallicity calibration deriving from direct abundance estimates could be devised based on the location of a galaxy along the $\rm O_{32}$ vs. $\rm R_{23}$ sequence. 

We investigate this claim further with \mygal\ and the comparison samples in the right panel of Figure \ref{fig:o32r23}. Here we have again plotted \mygal\ and the other samples on the high-excitation tail of the $\rm O_{32}$ vs. $\rm R_{23}$ diagram with each galaxy now color-coded by its direct metallicity estimate. Unlike in the left panel of Figure \ref{fig:o32r23}, we do not plot the error bars for the galaxies (except for \mygal) so as to more clearly illustrate any present trends. We see that there is indeed a nearly monotonic decrease in metallicity as one moves from the lower $\log(\rm O_{32})\sim$ 0.1 and $\log(\rm R_{23})\sim$ 0.8 along the sequence to higher values in both ratios. We also note that with redshift, there does not appear to be any significant evolution of the samples in either $\rm O_{32}$ or $\rm R_{23}$ as well as in metallicity. The $z\sim$ 0 sample from \citet{Izotov2006} and the $z\sim$ 0.8 sample from \citet{Jones2015} track the excitation sequence very similarly with comparable metallicity values as a function of position along the sequence. The intermediate- and high-redshift galaxies also do not collectively display any systematic offsets in their line-ratio values and do not show any evidence of evolution in their metallicities as a function of location on the sequence. These galaxies follow the same metallicity distribution seen by the lower-redshift samples. 

We do take note of the large scatter, particularly in $\log(\rm R_{23})$, of the $z>$ 1 galaxy sample. At fixed $\log(\rm O_{32})$, the galaxies of \citet[][C12]{Christensen2012} and \citet[][J14]{James2014} lie furthest to the left in $\log(\rm R_{23})$ compared to the lower-redshift samples while the galaxy of \citet[][S13]{Stark2013} and \mygal\ lie furthest to the right, having significantly higher $\rm R_{23}$ than the comparison samples. This observed scatter may be the consequence of underestimated uncertainties that do not account for systematic errors in the measurement and dust-correction of the emission lines, or it may hint at a larger intrinsic scatter in this line ratio at high redshift when compared to the relatively narrow high-excitation tail defined locally. In either case, our conclusions should not be significantly affected as $\rm R_{23}$, taken by itself, is not very sensitive to metallicity in the moderately sub-solar regime we are studying. A proper analysis of this scatter will require larger statistical samples with well constrained $\rm R_{23}$ and accurate metallicities that span a broad dynamic range.

The conclusions made from Figure \ref{fig:o32r23} support the findings of \citet{Shapley2015} of the $\rm O_{32}$ vs. $\rm R_{23}$ excitation diagram being a useful, redshift-invariant oxygen abundance indicator, based on the direct metallicity abundance scale, out to at least $z\sim$ 2.3 and perhaps $z\sim$ 3.1 with the inclusion here of COSMOS-1908 \citep{Sanders2016o3}. While much larger samples of intermediate- and high-redshift galaxies with direct metallicity estimates are required to confirm or refute the observed lack of evolution in this excitation diagram, its potential as an abundance indicator is important for several reasons \citep[see][]{Jones2015,Shapley2015,Sanders2016ne_u}. If this excitation sequence and its relation to metallicity are redshift-independent, then a local relation based on the much richer SDSS sample can be developed and applied accurately at high redshift. This sequence and a corresponding abundance calibration are based on line ratios solely involving strong oxygen and hydrogen emission lines, avoiding biases in nitrogen-based abundance indicators resulting from systematically higher N/O abundance ratios at high redshift \citep{Masters2014,Shapley2015,Sanders2016ne_u}. Finally, an indicator using this excitation sequence would be based on the direct metallicity abundance scale, with direct metallicities most closely reflecting the physical conditions present in star-forming regions due to their relation to electron temperature and density. 

\subsection{The Evolution of the Ionization Parameter}\label{subsec:ionparam}
The ionization parameter, defined as the ratio of the number density of hydrogen-ionizing photons to the number density of hydrogen atoms in the gas, characterizes the ionization state of the gas in a star-forming region and is often determined via the $\rm O_{32}$ (see Equation \ref{equ:o32}) line ratio. It has been suggested that at high redshift, galaxies have systematically higher ionization parameters than are usually found in local galaxies \citep{Brinchmann2008,Nakajima2013,Nakajima&Ouchi2014,Steidel2014,Kewley2015,Cullen2016,Kashino2017}. These studies have shown this largely based on comparisons at fixed stellar mass \citep[e.g.,][]{Kewley2015,Sanders2016ne_u}, comparison to the average ionization parameter of the entire SDSS \citep[e.g.,][]{Nakajima&Ouchi2014}, and comparisons at fixed metallicity \citep[e.g.,][]{Cullen2016,Kashino2017}. 

However, studying the [\ion{O}{3}]$\lambda$5007/[\ion{O}{2}]$\lambda\lambda$3726,3729 and [\ion{O}{3}]$\lambda$5007/H$\beta$ ratios at fixed metallicity in Figure \ref{fig:strong-lines}, we do not see any systematic offset of the high-redshift galaxies toward higher ionization parameter proxy (the former ratio) or higher excitation (the latter ratio) at fixed O/H. This is in agreement with \citet{Sanders2016o3}, who studied the same high-$z$ comparison galaxies, as well as \citet{Sanders2019}, who enlarged their high-$z$ sample with 3 new [\ion{O}{3}]$\lambda$4363 detections from the MOSDEF survey and \ion{O}{3}]$\lambda\lambda$1661,1666 sources from the literature. In regard to the former ratio, \mygal\ ($z$ = 2.59) and the $z$ = 1.43 galaxy of \citet{James2014} lie very close to the locally-calibrated, best-fit relation, within the 1$\sigma$ intrinsic scatter around the relation. The $z$ = 3.08 galaxy of \citet{Sanders2016o3} lies above the best-fit relation and scatter, but the $z$ = 1.43 galaxy of \citet{Stark2013} and the $z$ = 1.83 galaxy of \citet{Christensen2012} lie below them. When considering the latter ratio, all four high-redshift galaxies lie near the best-fit relation within the intrinsic scatter. These results from Figure \ref{fig:strong-lines} are corroborated in the $\rm O_{32}$ vs. $\rm R_{23}$ excitation diagram of Figure \ref{fig:o32r23}. We see no collective systematic offset of these galaxies in $\rm O_{32}$ at fixed $\rm R_{23}$ (a diagnostic for oxygen abundance).

The conclusions drawn from Figures \ref{fig:strong-lines} and \ref{fig:o32r23} contrast with studies such as \citet{Cullen2016} and \citet{Kashino2017}, who argue for increased ionization parameter at fixed O/H in high-redshift galaxies. Instead, our results support the suggestions of \citet{Sanders2016o3,Sanders2016ne_u,Sanders2019}, who argue for an absence of evolution in the ionization parameter at fixed metallicity. \citet{Sanders2016ne_u} used $\sim100$ star-forming galaxies at $z\sim$ 2.3 from the MOSDEF survey to suggest that while high-redshift galaxies do in fact have systematically higher $\rm O_{32}$ values at fixed stellar mass relative to local galaxies, they have similar $\rm O_{32}$ values at fixed $\rm R_{23}$. They argue that, with the high-redshift MOSDEF sample following the same distribution as local galaxies along the higher $\rm O_{32}$ and $\rm R_{23}$ end of the excitation sequence, and this end corresponding to lower metallicities \citep{Shapley2015}, the ionization state of high-redshift, star-forming galaxies must be similar to metal-poor local galaxies. This is corroborated by \citet{Sanders2019}, who show that, on average, their $z>$ 1 auroral-line-emitting sample lies on local relations between ionization parameter and direct-method oxygen abundance, positioned in the same location as metal-poor, $z\sim$ 0 SDSS stacks and local \ion{H}{2} regions. \citet{Sanders2016ne_u} further argue that the difference in offset when comparing to constant stellar mass as opposed to constant metallicity is due to the evolution of the mass-metallicity relation, where high-redshift galaxies have systematically lower metallicities than local galaxies at fixed stellar mass \citep{Sanders2015}.

It is important to note that the results of this paper support the notion of a lack of evolution in ionization parameter at fixed metallicity without the use of nitrogen in the metallicity estimates. As stated earlier, using direct metallicities and diagnostics ($\rm R_{23}$) not involving nitrogen avoids possible systematic offsets in the abundance estimates due to higher N/O abundance ratios at high redshift.

\subsection{Low-Mass End of the Fundamental Metallicity Relation}\label{subsec:fmr}
The Fundamental Metallicity Relation \citep{Mannucci2010} is a 3D surface defined by a tight dependence of gas-phase metallicity on stellar mass and SFR and is suggested to exist from $z$ = 0 out to $z$ = 2.5 without evolution \citep[e.g.,][]{Mannucci2010,Henry2013,Maiolino&Mannucci2019}. From this surface, \citet{Mannucci2010} define a projection, $\mu_\alpha$ vs. 12+$\log$(O/H), where $\mu_\alpha$ is a linear combination of stellar mass and SFR relying on the observed correlation and anti-correlation of metallicity with stellar mass and SFR, respectively.  

\begin{equation}\label{equ:mu32}
    \mu_\alpha\ =\ \log(M_\ast) - \alpha\log(\rm SFR)
\end{equation}

\noindent \citet{Mannucci2010} suggest that if $\alpha$ = 0.32 in this relation, the scatter in metallicity at fixed $\mu_\alpha$ is minimized, all galaxies out to $z$ = 2.5 show the same dependence of metallicity on $\mu_{0.32}$, and all galaxies out to this redshift occupy the same range of $\mu_{0.32}$ values.

Unfortunately, the FMR of \citet{Mannucci2010} is defined by low-redshift SDSS galaxies with stellar masses down to $\log(M_\ast/M_\sun)$ = 9.2, $\sim$1.1 (0.6) dex above the lower- (upper-) limit stellar mass of \mygal\ (see Section \ref{subsec:stellarmass} and Figure \ref{fig:sed}). In SFR, this FMR only probes galaxies with $-1.45\leqslant$ log(SFR) $\leqslant0.8$, whereas \mygal\ has a log(SFR) = 1.2. Furthermore, the redshift-invariant nature of the FMR and $\mu_{0.32}$ $-$ metallicity projection only applies out to $z$ = 2.5, with \mygal\ lying just beyond this redshift at $z$ = 2.59. Perhaps most importantly, the \citet{Mannucci2010} FMR is defined with metallicities calculated via locally-calibrated, strong-line diagnostics \citep{Maiolino2008}, the applicability of such indirect methods at high redshift being a primary focus of this paper.

\begin{figure*}[ht!]
    \includegraphics[trim={0cm 0cm 0.5cm 0.4cm}, width=1.05\columnwidth, clip]{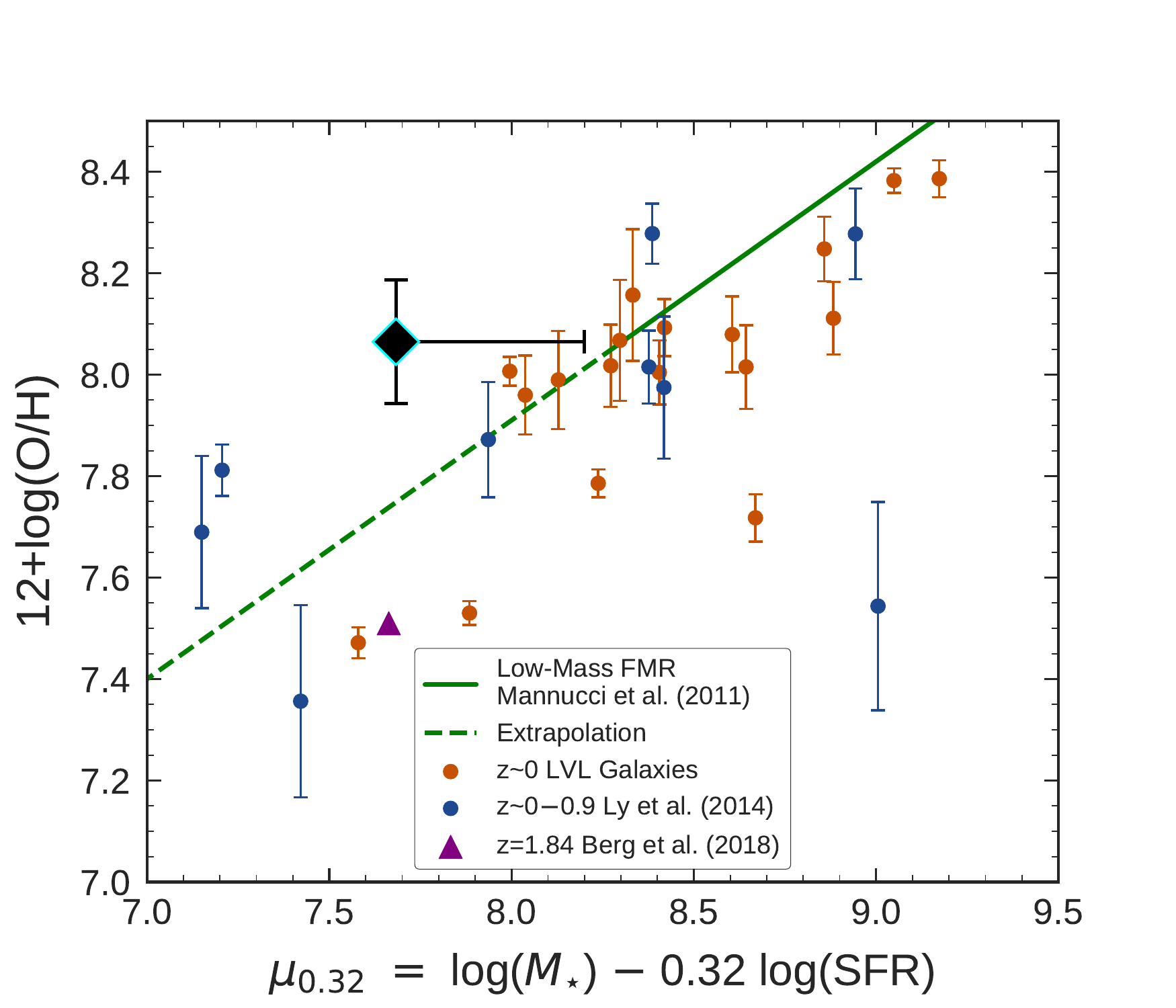}\quad\includegraphics[trim={0cm 0cm 0.5cm 0.4cm}, width=1.05\columnwidth, clip]{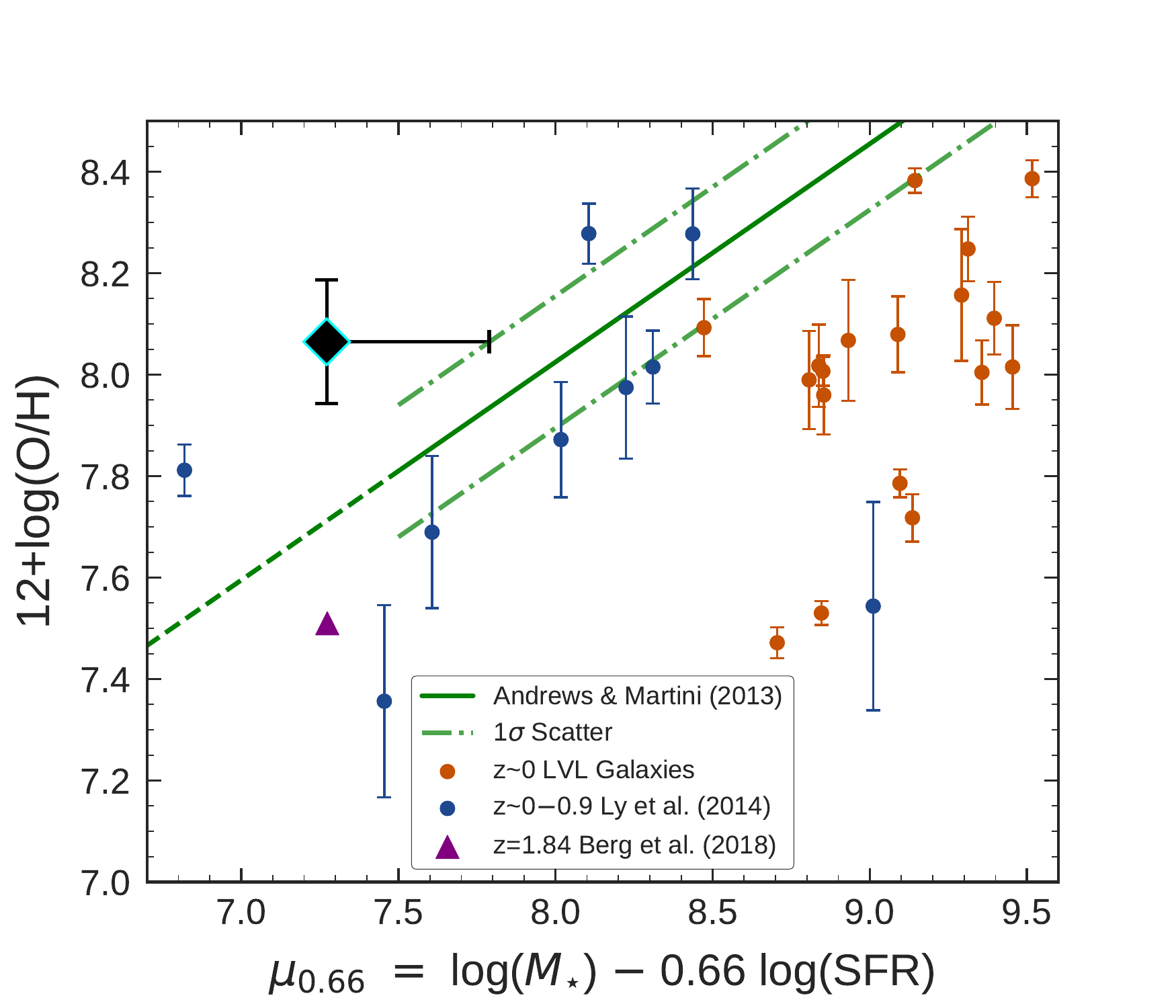}
    \caption{(Left) The low-mass extension of the Fundamental Metallicity Relation \citep[FMR;][]{Mannucci2010} as given by the projection of metallicity vs. $\mu_\alpha$ (in solar units; see Equation \ref{equ:mu32}) for $\alpha$ = 0.32. This extension (Equation \ref{equ:fmr}) was calculated by \citet{Mannucci2011} down to $\mu_{0.32}\sim$ 8.3 (solid line), so an extrapolation in $\mu_{0.32}$ is shown here for lower values (dashed line). (Right) The FMR, metallicity vs. $\mu_\alpha$ projection as calculated by \citet{Andrews&Martini2013}. These authors report a minimization in the scatter of metallicity at fixed $\mu_\alpha$ for $\alpha$ = 0.66. This linear relation, with slope $m$ = 0.43, is only calibrated down to $\mu_{0.66}\sim$ 7.5 (solid line), so an extrapolation in $\mu_{0.66}$ is given here (dashed line). The scatter in the projection (based on stacks instead of individual galaxies) is given to be $\sigma$ = 0.13 dex and is shown by the dot-dashed lines. Metallicities used in \citet{Mannucci2010,Mannucci2011} are based on strong-line methods whereas metallicities in \citet{Andrews&Martini2013} are $T_e$-based from stacks of SDSS spectra. In both panels, \mygal\ is given by the black diamond with cyan border. Its stellar mass of $\log$($M_\ast/M_\sun$) = 8.07 is likely a lower limit not accounting for an undetected older stellar population (see Section \ref{subsec:stellarmass} and Figure \ref{fig:sed}), so we show the increase \mygal\ would experience in $\mu_{\alpha}$ for a factor of $\sim$ 3.3 increase in stellar mass. A similar galaxy (in $M_\ast$ and SFR) to \mygal\ from \citet{Brammer2012} and \citet{Berg2018} has its lower-limit metallicity (see text for details) plotted as the purple triangle. A sample of $z<0.9$ galaxies from \citet{Ly2014} is shown in blue, and a low-mass, $z\sim$ 0, Local Volume Legacy (LVL) sample (see text for details) is shown by the dark orange data points. All galaxy samples have their direct metallicities plotted in both panels.\label{fig:fmr}}
\end{figure*}

Addressing the limited stellar mass range, \citet{Mannucci2011} extended the FMR, or more specifically the $\mu_{0.32}$ $-$ metallicity projection, down to a stellar mass of $\sim \rm 10^{8.3}\ M_\sun$ using $\sim$ 1300 galaxies from the \citet{Mannucci2010} sample with 8.3 $<$ $\mu_{0.32}$ $<$ 9.4. They found that these low-mass galaxies extend the FMR with a smooth, linear relation between gas-phase metallicity and $\mu_{0.32}$ given, for $\mu_{0.32}$ $<$ 9.5, by:

\begin{equation}\label{equ:fmr}
    12+\log(\rm O/H)\ =\ \rm 8.93+0.51(\mu_{\rm 0.32}-10)
\end{equation}

Recognizing that metallicity estimates based on different methods can differ drastically for the same galaxies \citep{Kewley&Ellison2008}, \citet{Andrews&Martini2013} investigated the $\mu_\alpha$ (Equation \ref{equ:mu32}) FMR projection using the $T_e$-based metallicities they calculated with their stacked SDSS spectra. Using galaxies with 7.5 $\lesssim$ log($M_\ast/M_\sun$) $\lesssim$ 10.6 and $-1.0$ $\leqslant$ log(SFR) $\leqslant$ 2.0 binned by $M_\ast$ and SFR, they found that $\alpha$ = 0.66 minimized the scatter in their metallicities at fixed $\mu_\alpha$. While this calibration of the $\mu_\alpha$ $-$ metallicity projection utilizes direct-method oxygen abundances, it still suffers from both a lack of high-redshift data due to the faintness of $T_e$-sensitive auroral lines and a poor sampling of low-mass, high-SFR galaxies like \mygal\ (see Figure 1 of \citealt{Andrews&Martini2013} for the distribution in $M_\ast$ and SFR of their sample).

We test the validity of the FMRs of \citet{Mannucci2011} and \citet{Andrews&Martini2013} in the poorly-sampled $M_\ast-\rm SFR$ parameter space occupied by \mygal. In Figure \ref{fig:fmr}, we plot \mygal\ against the low-mass FMR extension (left) given by Equation \ref{equ:fmr}, extrapolated down by $\sim$ 0.6 dex in $\mu_{0.32}$, and against the $T_e$-based FMR (right), extrapolated down by $\sim$ 0.2 dex in $\mu_{0.66}$. We also plot the $z$ = 1.84 highly-ionized, lensed galaxy (SL2SJ02176-0513) of \citet{Brammer2012} and \citet{Berg2018}, which, when adjusted for a \citet{Chabrier2003} IMF with 0.2 $\rm Z_\sun$, has a very similar stellar mass (log($M_\ast/M_\sun$) = 8.03) and SFR (14 $\rm M_\sun\ yr^{-1}$) as \mygal. Despite these similar properties, SL2SJ02176-0513 has a much lower metallicity (12+log(O/H) $\geqslant$ 7.51) than \mygal, however. We note that its metallicity is reported as a lower limit due to both the lack of spectroscopic coverage of the [\ion{O}{2}]$\lambda\lambda$3726,3729 emission lines needed for the determination of $\rm O^+/H^+$ (see Equation \ref{equ:o/h}) and the possibility of a contribution from $\rm O^{+3}$ to O/H. Nevertheless, as detailed in \citet{Berg2018}, this lower limit should be close to the actual value as the highly-ionized nature of the galaxy makes the $\rm O^+$ contribution to the oxygen abundance very small (estimated at 2$\%$ of the total oxygen abundance; included in our stated lower-limit metallicity), and the ionization correction factor (ICF) for contribution of $\rm O^{+3}$ is also estimated to be small (ICF = 1.055; not included in our stated lower-limit metallicity).

For further comparison of \mygal\ and the FMRs to other low-mass galaxies spanning a broad range of star formation activity, we also include in Figure \ref{fig:fmr} the partial \citet{Ly2014} sample used in this work (median log($M_\ast/M_\sun$) $\sim$ 8.4 and median specific star formation rate (sSFR) $\sim$ 9.3 $\rm Gyr^{-1}$) and a $z\sim$ 0 LVL subsample (median log($M_\ast/M_\sun$) $\sim$ 7.7 and median sSFR $\sim$ 0.2 $\rm Gyr^{-1}$). The \citet{Ly2014} sample, in addition to using the metallicities re-derived in this work, uses SFRs re-calculated assuming a \citet{Cardelli1989} extinction law. Stellar masses for this sample are the values given in \citet{Ly2014} for a \citet{Chabrier2003} IMF with 0.2 $\rm Z_\sun$. The LVL objects used here comprise the subset of the \citet{Berg2012} sample used in Figures \ref{fig:bpt} and \ref{fig:strong-lines} of which the objects are a part of both the sample used in \citet{Berg2012} and the sample in \citet{Weisz2012}. Metallicities used here are those re-calculated in this paper with the emission-line fluxes from \citet{Berg2012}. Stellar masses for these galaxies are taken from \citet{Weisz2012} while the SFRs are calculated from H$\alpha$ measurements taken by \citet{Kennicutt2008} and \citet{Lee2009} as part of the 11HUGS survey. All SFRs for \mygal\ and the comparison samples are calculated via Balmer recombination lines, assuming a \citet{Chabrier2003} IMF with 0.2 $\rm Z_\sun$, and all metallicities are calculated via the ``direct" method.

With the lower-limit stellar mass estimated by our SED fitting ($\log$($M_\ast/M_\sun$) = 8.07), \mygal\ lies $\sim$ 2.6$\sigma$ (2.9$\sigma$) above the extrapolation of the low-mass FMR extension of \citet{Mannucci2011} ($T_e$-based FMR of \citealt{Andrews&Martini2013}). However, as mentioned in Section \ref{subsec:stellarmass} and seen in Figure \ref{fig:sed}, an unseen, older stellar population component can exist in \mygal\ without significantly altering the observed SED, raising the stellar mass estimate of \mygal\ by as much as a factor of 3.3 (up to $\log$($M_\ast/M_\sun$) = 8.59). An increase in stellar mass will correspondingly increase the measured value of $\mu_\alpha$ (Equation \ref{equ:mu32}) and bring \mygal\ into better agreement with both FMRs. This is seen in Figure \ref{fig:fmr}, where the horizontal bar extending from \mygal\ represents the range of $\mu_\alpha$ values corresponding to our estimated range of stellar masses for \mygal. If the mass estimate is even $\sim$ 2$\times$ what we state as the lower bound, \mygal\ is consistent with the FMR of \citet{Andrews&Martini2013} within the 1$\sigma$ scatter around the relation and the uncertainty in \mygal's oxygen abundance. Without this mass increase, \mygal\ is very likely already consistent with the extrapolation of the low-mass end of the FMR as given by \citet{Mannucci2011} considering the 1$\sigma$ dispersions in metallicity seen at fixed $\mu_{0.32}$ in their work (see right panel of their Figure 1). We therefore suggest that \mygal\ is consistent with both FMRs within the observed scatter around each relation.

An important takeaway from Figure \ref{fig:fmr} is the large scatter seen around both $\mu_\alpha$ $-$ metallicity projections. This is well illustrated when comparing \mygal\ and SL2SJ02176-0513 from \citet{Berg2018}. Despite having similar $\rm sSFRs \sim 135\ Gyr^{-1}$, these galaxies differ in oxygen abundance by $\sim$ 0.55 dex, lying on either side of both FMRs. Large scatter is also seen in the \citet{Ly2014} comparison sample, despite the sample being generally consistent with both FMRs. This scatter observed in Figure \ref{fig:fmr} around the FMRs is likely due to the increased variation in star formation histories and current star formation activity in dwarf galaxies \citep{Mannucci2011,Emami2018} and suggests that physical processes of gas flows, enrichment, and star formation have not yet reached equilibrium \citep{Ly2015}. Physical timescale effects in dwarf galaxies with bursty star formation may lead to large dispersions in the metallicities of galaxies with similar properties, like we see with \mygal\ and SL2SJ02176-0513, whereby we may be observing more metal-rich galaxies at a time when recent star formation has enriched the gas, but not yet removed metals from the galaxy via supernovae and other stellar feedback \citep{Ly2015}. 

In consideration of the LVL sample here, we note the systematic offsets of the galaxies (median log(SFR) $\sim -1.9$ and median $\log(M_\ast/M_\sun)\sim$ 7.7) particularly from the relation of \citet{Andrews&Martini2013}, but also slightly below the relation of \citet{Mannucci2011} on average. While an in-depth study of these offsets is beyond the scope of this work, they may arise from a lack of examination of the $M_\ast \sbond$ SFR parameter space occupied by the LVL galaxies. \citet{Mannucci2011} only probe down to $M_\ast \sim 10^{8.3}\ \rm M_\sun$ and $\log$(SFR) $\sim -1.45$ while \citet{Andrews&Martini2013} study a sample with the vast majority of objects having $\log$(SFR) $> -1$ and $\log(M_\ast/M_\sun)$ $>$ 8. The extreme offset of the LVL galaxies from the \citet{Andrews&Martini2013} relation may also result from the stronger dependence of $\mu_\alpha$ on SFR in this calibration ($\alpha$ = 0.66) compared to that in \citet{Mannucci2010} ($\alpha$ = 0.32).

\subsection{A Comparison Against the MZR Predictions of FIRE}\label{subsec:FIRE}
The FIRE\footnote{\url{https://fire.northwestern.edu/}} (Feedback In Realistic Environments) simulations \citep{Hopkins2014} are cosmological zoom-in simulations that contain realistic physical models and resolution of the multi-phase structure of the ISM, star formation, and stellar feedback. \citet{Ma2016} utilize these simulations to study the evolution of the stellar mass $-$ gas-phase metallicity relation from $z=0-6$ for galaxies spanning the stellar mass range $M_\ast=10^4-10^{11}\ \rm M_\sun$ at $z=0$. They predict an MZR that has a slope which does not vary appreciably with redshift. They fix the slope to the mean value with redshift, $m=0.35$ (which almost perfectly agrees with the best-fit slope between $z=1.4$ and $z=3.0$ $\sbond$ see their Figure 3), and report an MZR that evolves with $z$ as:

\begin{eqnarray}\label{equ:FIRE}
    12 + \log(\rm O/H)\ =\ && 0.35[\log(M_\ast/M_\sun) - 10] \nonumber \\
    && + 0.93\exp(-0.43z) + 7.95
\end{eqnarray}

Comparing \mygal\ against this prediction, at $z=\redshift$, with \mygal's lower- (upper-) limit stellar mass of $\log(M_\ast/M_\sun)=8.07$ (8.59; see Section \ref{subsec:stellarmass} and Figure \ref{fig:sed}), we find that the metallicity of \mygal\ ($12+\log(\rm O/H)=8.06\pm0.12$) is $\sim 4.0\sigma$ (2.5$\sigma$) above the predicted oxygen abundance of $12+\log(\rm O/H)=7.58$ (7.76). Comparing the prediction in Equation \ref{equ:FIRE} also against the galaxy, SL2SJ02176-0513, of \citet{Berg2018} at $z=1.8444$ and $\log(M_\ast/M_\sun)=8.03$, we find that the lower-limit metallicity of the galaxy (7.51; see \citealt{Berg2018} and Section \ref{subsec:fmr} for details on the lower limit) lies 0.17 dex below the prediction of $12+\log(\rm O/H)=7.68$. Further comparing the position of both of these galaxies to the scatter around the MZR in Figure 3 of \citet{Ma2016}, we see that \mygal\ lies above all simulated galaxies at its lower-limit stellar mass, but likely among the objects scattered high in oxygen abundance at its upper-limit stellar mass. SL2SJ02176-0513 lies below the best-fit relation, but is consistent within the scatter. 

Considered together, despite being at different redshifts, these results at least show that there is significant scatter of dwarf galaxies around the MZR at roughly fixed stellar mass. This is likely due to time variations in the metallicities of dwarf galaxies resulting from the bursty nature of their star formation and its connection to gas inflows/outflows \citep{Ma2016}. Due to the extremely metal-poor nature of SL2SJ02176-0513 ($\sim0.07$ $\rm Z_\sun$) and its general agreement with the predicted MZR, as well as the discrepancy of \mygal\ from the MZR, particularly when considering the lower-end of \mygal's mass range, these results may also suggest that the slope ($m=0.35$) in Equation \ref{equ:FIRE} is too steep. However, larger observational samples are needed to verify this suggestion.

\section{Summary}\label{sec:summary}
In this paper, we present a 4.2$\sigma$ detection of the temperature-sensitive, auroral [\ion{O}{3}]$\lambda$4363 emission line in a lensed, star-forming, dwarf galaxy at $z$ = 2.59, \mygal. With the extinction-corrected fluxes of the rest-optical, nebular emission lines, we estimate the electron temperature and density of this galaxy and calculate, directly, an oxygen abundance of 12+$\log$(O/H) = 8.06 $\pm$ 0.12 (0.24 $\rm Z_\sun$). With this measurement, and intrinsic strong-line ratios calculated for \mygal, we report the following:
\begin{enumerate}
    \item We study the evolution with redshift of strong-line ratio $-$ direct metallicity relations calibrated and suggested to be redshift-invariant out to $z\sim0.8$ by \citet{Jones2015}. With a $z\sim0$ comparison sample from \citet{Izotov2006}, the 32 $z\sim0.8$ galaxies from \citet{Jones2015}, 9 $z<$ 0.9 galaxies from \citet{Ly2014}, and 4 high-redshift galaxies ($z$ = 1.43, 1.43, 1.83, 3.08) with [\ion{O}{3}]$\lambda$4363 detections in addition to \mygal, we find no evidence for evolution of the Jones et al. strong-line ratio $-$ metallicity calibrations. We also study the [\ion{O}{3}]/[\ion{O}{2}] metallicity calibration of \citet{Bian2018}, the preferred metallicity diagnostic in the strong-line metallicity study of \citet{Sanders2019}. We find general agreement between this relation and our samples as well as with the relation of \citet{Jones2015}. We note divergences from the Jones et al. relations of our $z\sim$ 0 LVL \ion{H}{2} region sample below $12+\log(\rm O/H)\sim$ 8.1, similar to \ion{H}{2} region divergences seen in \citet{Sanders2019}.
    \item Using the same comparison samples, we find no significant evolution with redshift in the high-excitation tail of the $\rm O_{32}$ vs. $\rm R_{23}$ excitation diagram. The different galaxy samples do not display any relative offsets in either $\rm O_{32}$ or $\rm R_{23}$, with intermediate- and high-redshift galaxies following the same distribution as local galaxies, albeit with larger scatter of the $z>1$ sample in $\log(\rm R_{23})$. We also observe the nearly monotonic decrease in direct metallicity with increasing $\rm O_{32}$ and $\rm R_{23}$ seen in \citet{Shapley2015}. As with the strong-line ratios, we find no evidence for evolution with redshift of the metallicity as a function of position along the excitation sequence. The combination of these results supports the conclusions of \citet{Shapley2015} that the $\rm O_{32}$ vs. $\rm R_{23}$ excitation diagram can be a useful, direct-metallicity-based, redshift-invariant, empirical oxygen abundance indicator.
    \item Through our study of both the strong-line ratio $-$ metallicity relations and the $\rm O_{32}$ vs. $\rm R_{23}$ excitation diagram, we find no evolution with redshift of the ionization parameter at fixed O/H. This result is in agreement with \citet{Sanders2016o3,Sanders2016ne_u,Sanders2019}, who report the same finding and suggest that the ionization state of high-$z$, star-forming galaxies is similar to local, metal-poor galaxies.
    \item We plot \mygal\ against both the $\mu_{0.32}$ $-$ metallicity projection of the Fundamental Metallicity Relation (FMR) as extended to low stellar mass by \citet{Mannucci2011} and the $\mu_{0.66}$ $-$ metallicity projection of \citet{Andrews&Martini2013}, wherein the metallicities are $T_e$-based as opposed to the strong-line basis of \citet{Mannucci2011}. The stated stellar mass range ($\log$($M_\ast/M_\sun$) = $8.07-8.59$) and SFR (16.2 $\rm M_\sun\ yr^{-1}$) of \mygal\ yield a range in $\mu_{0.32}$ ($\mu_{0.66}$) of $\sim7.7-8.2$ ($\sim7.3-7.8$) and thus require slight extrapolations of both FMRs in $\mu_\alpha$ ($\sim0.6$ dex in $\mu_{0.32}$ and $\sim0.2$ dex in $\mu_{0.66}$). We also compare \mygal\ and the FMRs to other low-mass galaxy samples at low-to-high redshift with a large range in current star formation activity. Together, these samples show a large scatter around the FMR, likely due to large variations in star formation history and current star formation activity in dwarf galaxies. With this observed scatter, and the uncertain mass estimate of \mygal\ resulting from the possibility of the presence of an unseen, older stellar population within the galaxy, we conclude that \mygal\ is consistent with both FMRs studied.
    \item We compare the locations in $M_\ast-Z$ parameter space of \mygal\ and the galaxy from \citet{Berg2018} to the predicted MZR from the FIRE hydrodynamical simulations \citep{Ma2016}. \mygal\ lies $\sim0.3-0.5$ dex above the predicted relation while the object from \citet{Berg2018} lies $\sim0.2$ dex below the relation, suggesting a large scatter in the relation at low-mass and/or a slightly shallower MZR slope than predicted.
\end{enumerate}

This study adds another crucial data point at high redshift in terms of direct oxygen abundance estimates and dwarf galaxy properties. With the measurements of \mygal\ and their comparisons to measurements of other auroral-line-emitting galaxies at various redshifts, we are able to further constrain the validity of several diagnostics at high redshift and low stellar mass, such as locally-calibrated strong-line ratio $-$ direct metallicity relations and the FMR. However, large statistical samples of high-redshift [\ion{O}{3}]$\lambda$4363 sources and very low mass dwarf galaxies are needed to properly constrain these diagnostics. Regardless, this and other similar studies help to prepare us for those large surveys that will be conducted with the next generation of ground and space-based telescopes.

\acknowledgments

This material is based upon work supported by the National Science Foundation under Grant No. 1617013. 

Support for programs $\#$12201 and $\#$12931 was provided by NASA through a grant from the Space Telescope Science Institute, which is operated by the Association of Universities for Research in Astronomy, Inc., under NASA contract NAS5-26555.

The authors wish to recognize and acknowledge the very significant cultural role and reverence that the summit of Maunakea has always had within the indigenous Hawaiian community.  We are most fortunate to have the opportunity to conduct observations from this mountain.

\vspace{6mm}
\facilities{Keck:I (MOSFIRE, LRIS), HST (WFC3, ACS)}
\vspace{6mm}
\appendix 
\vspace{-3cm}
\section{Yuan 2009 Detection}\label{sec:appendix}

This paper includes a re-analysis of previously reported high-redshift ($z>1$) detections of [\ion{O}{3}]$\lambda$4363. \citet{Yuan&Kewley2009} reported a $\sim3\sigma$ detection of [\ion{O}{3}]$\lambda$4363 in a $z=1.7$ galaxy behind Abell 1689, referred to as ``Lens22.3'' in their paper and first reported as a multiply-imaged galaxy in \citet{Broadhurst2005}. As part of our larger campaign to obtain near-IR spectra of lensed, high-redshift galaxies, we obtained a MOSFIRE J-band spectrum of Lens22.3 as well as of another image of the same galaxy \citep[referred to as Lens22.1 in][]{Broadhurst2005}. Both images were observed in the same slit mask for 1,440 seconds on 2015 January 20 and 4,320 seconds on 2016 February 1 in $\sim0\farcs6$ seeing on both nights. Though our exposure times are somewhat shorter than the \citet{Yuan&Kewley2009} observations (5,760 seconds vs. 6,800 seconds), the much higher spectral resolution ($R\sim3300$ vs. $R\sim500$) and narrower slit width (0\farcs7 vs. 1\farcs0) of the MOSFIRE observations result in a superior sensitivity to narrow emission lines. For a specific comparison in the J-band, our detections of H$\beta$ are $35\sigma$ and $28\sigma$ for Lens22.3 and Lens22.1, respectively, compared to $10\sigma$ for the \citet{Yuan&Kewley2009} detection. For additional sensitivity to faint lines, we normalized the two spectra (by the [\ion{O}{3}]$\lambda$4959 flux) and created a weighted-average spectrum, resulting in an H$\beta$ detection of $48\sigma$.  

\begin{figure*}[ht!]
    \includegraphics[trim={00.2cm 0.65cm 1.cm 1cm}, width=\columnwidth, clip]{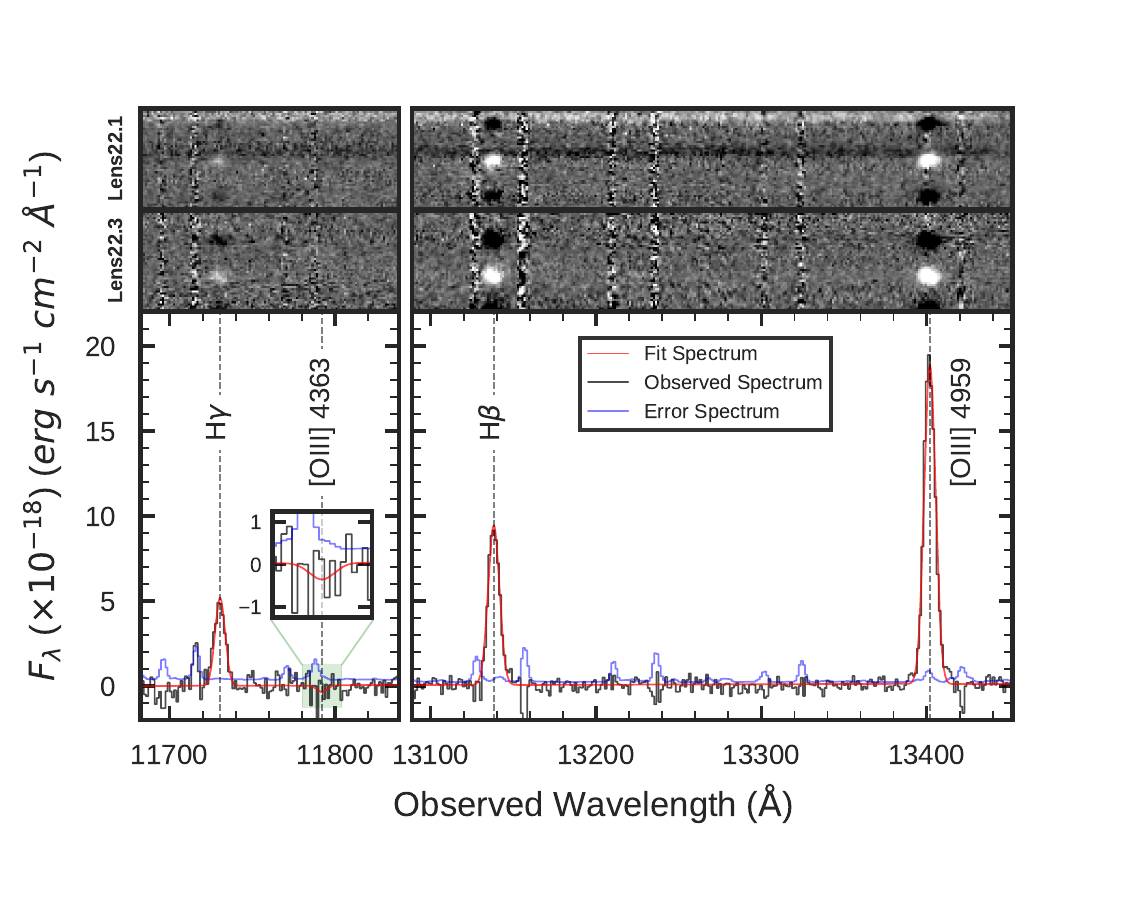}
    \caption{The 2D spectra of Lens22.1 (top) and Lens22.3 (bottom) (images referred to in \citealt{Broadhurst2005}), two images of the same galaxy at $z=1.7026$. Below is plotted the combined 1D spectrum of both images (black), the error spectrum (blue), and the best-fit continuum and emission lines (red). Strong emission lines are seen in H$\gamma$, H$\beta$, and [\ion{O}{3}]$\lambda$4959, but no detection is seen in [\ion{O}{3}]$\lambda$4363 (either in 1D or 2D), in disagreement with the claimed detection in \citet{Yuan&Kewley2009}. The portion of the 1D spectrum containing [\ion{O}{3}]$\lambda$4363 has been highlighted in green and magnified in the inset plot.}
    \label{fig:370_1197}
\end{figure*}

The 2D spectra of Lens22.3 and Lens22.1 and the stacked 1D spectrum can be seen in Figure \ref{fig:370_1197}. Strong  [\ion{O}{3}]$\lambda$4959, H$\beta$, and a $23\sigma$ detection of $H\gamma$ can be seen.  However, there is no evidence of an [\ion{O}{3}]$\lambda$4363 line. Given the reported $H\beta$/[\ion{O}{3}]$\lambda$4363 $\sim3.7$, we should have detected the line at $\sim9.2\sigma$. Given the much lower spectral resolution of the Subaru/MOIRCS spectrum of \citet{Yuan&Kewley2009}, we believe that the line detected in the MOIRCS spectrum was likely the H$\gamma$ line. That would also help explain why the line center reported in that spectrum was at a somewhat lower redshift than the other lines ($z=1.696$ vs. $z=1.705$).

A more detailed analysis of this spectrum and the rest of our sample will be reported in future works. 

\vspace{-0.25cm}

\begin{deluxetable}{lrr}[ht]
\tablecaption{Emission-Line Fluxes of Lens22.3 and Lens22.1 \label{tab:370_1197}}
\tablecolumns{3}
\tablenum{3}
\setlength{\tabcolsep}{5pt}
\tablewidth{\textwidth}
\tablehead{
\colhead{Line} &
\colhead{Relative Flux\tablenotemark{a}} &
\colhead{S/N}
}
\startdata
H$\gamma$ & 0.49 & 23 \\
$[$\ion{O}{3}$]\lambda$4363 & $<0.03$ & \nodata \\
H$\beta$ & 1.0 & 48 \\
$[$\ion{O}{3}$]\lambda$4959 & $2.00$ & 67 \\
\enddata
\tablenotetext{a}{Fluxes relative to H$\beta$ flux}
\end{deluxetable}

\vspace{-1.95cm}

\pagebreak

\bibliography{id217.bib}{}
\bibliographystyle{aasjournal.bst}

\end{document}